\documentclass[preprint]{aastex}

\newcommand{\ltwid}{\mathrel{\raise.3ex\hbox{$<$\kern-.75em\lower1ex\hbox{$\sim$
}}}}
\newcommand{\gtwid}{\mathrel{\raise.3ex\hbox{$>$\kern-.75em\lower1ex\hbox{$\sim$
}}}}
\newcommand{\bd}{\begin{description}}
\newcommand{\ed}{\end{description}}
\newcommand{\p}{\s\prime}

\newcommand{\s}{\scriptscriptstyle}
\begin{document}

\title{Complete Tidal Evolution of Pluto-Charon}
\author{W.~H.~Cheng$^{\rm a}$, Man Hoi Lee$^{\rm a,b}$, S.~J.~Peale$^{\rm c}$}
\affil{$^{\rm a}$Department of Earth Sciences, The University of Hong Kong,
  Pokfulam Road, Hong Kong}
\affil{$^{\rm b}$Department of Physics, The University of Hong Kong,
  Pokfulam Road, Hong Kong}
\affil{$^{\rm c}$Department of Physics, University of California,
  Santa Barbara, CA 93106}

\begin{abstract}
Both Pluto and its satellite Charon have rotation rates synchronous
with their orbital mean motion.  This is the theoretical end point of
tidal evolution where transfer of angular momentum has ceased.  Here
we follow Pluto's tidal evolution from an initial state having the
current total angular momentum of the system but with Charon in an
eccentric orbit with semimajor axis $a \approx 4R_{\s P}$ (where
$R_{\s P}$ is the radius of Pluto), consistent with its impact origin.
Two tidal models are used, where the tidal dissipation function
$Q\propto$ 1/frequency and $Q=$ constant, where details of the
evolution are strongly model dependent. The inclusion of the
gravitational harmonic coefficient $C_{22}$ of both bodies in the
analysis allows smooth, self consistent evolution to the dual
synchronous state, whereas its omission frustrates successful
evolution in some cases.  The zonal harmonic $J_2$ can also be
included, but does not cause a significant effect on the overall
evolution.  The ratio of dissipation in Charon to that in Pluto
controls the behavior of the orbital eccentricity, where a judicious
choice leads to a nearly constant eccentricity until the final
approach to dual synchronous rotation. The tidal models are complete
in the sense that every nuance of tidal evolution is realized while
conserving total angular momentum --- including temporary capture into
spin-orbit resonances as Charon's spin decreases and damped librations
about the same.
\end{abstract}

\section{INTRODUCTION}

Pluto has five known satellites: Charon, Nix, Hydra, Keberos, and
Styx, with the latter four much smaller than Charon.
Listed in Table~\ref{para} are the physical and orbital parameters of
Pluto-Charon from \cite{Buie12}, unless otherwise specified.
The Charon-Pluto mass ratio ($q=0.1165$) is large when compared with
others in the Solar System ($1/81$ for Moon-Earth and $< 1/4000$ for
the other satellites and their planets). The barycenter of the
Pluto-Charon system lies outside the surface of Pluto. Hence, some
astronomers regard the pair as a binary system \citep{Stern92}. The
total angular momentum $L$ of the Pluto-Charon system is so large that
the combined pair would be rotationally unstable
\citep{Mignard81a,Lin81}.

The Pluto-Charon system is currently in a dual synchronous state
\citep{Buie97,Buie10},
which is the endpoint of tidal evolution. As such the expected zero
orbital eccentricity has been recently verified (with a 1-$\sigma$
upper limit of $7.5 \times 10^{-5}$), after taking into account the
effects of surface albedo variations on Pluto (\citealt{Buie12}; see
Table~\ref{para}).

\begin{deluxetable}{lcc}
\tablecolumns{3}
\tablewidth{0pt}
\tablecaption{Physical and orbital parameters of the Pluto-Charon system
\label{para}}
\tablehead{
\colhead{Parameter} & \colhead{Pluto} & \colhead{Charon}
}
\startdata
$GM$ (km$^3$\,s$^{-2}$)\tablenotemark{a} & 870.3(3.7) & 101.4(2.8) \\
Radius $R$ (km)\tablenotemark{b} & 1153(10) & 606.0(1.5) \\
Orbital period $P$ (days) & \multicolumn{2}{c}{6.3872273(3)} \\
Semimajor axis $a$ (km) & \multicolumn{2}{c}{19573(2)} \\
Eccentricity $e$ &\multicolumn{2}{c}{0} \\
Inclination $i$ ($^{\circ}$)&\multicolumn{2}{c}{96.218(8)} \\
Long.\ ascending node $\Omega$ ($^{\circ}$) & \multicolumn{2}{c}{223.0232(69)} \\
\enddata
\tablenotetext{a}{Adopted from \cite{Tholen08}, where $G$ is the
  Newtonian gravitational constant and $M$ is the mass.}
\tablenotetext{b}{Adopted from \citet{Buie06} for Pluto and
  \citet{Person06} for Charon.}
\tablecomments{
The orbital elements are Pluto-centric with respect to the
mean equator and equinox of J2000 at the epoch JD 2452600.5.
Numbers in parentheses are 1-$\sigma$ errors in the least significant
digits.} \label{parameters}
\end{deluxetable}

As Pluto-Charon is similar to Earth-Moon, the feasible origin of this
system may be chosen from the proposed schemes for the origin of the
Earth-Moon system.  A giant impact of a Mars-sized body is thought to
be the only viable origin of the Moon  (e.g.,
\citealt{Cameron76,Boss86,Canup04}) to account for the large angular
momentum of the system. \cite{McKinnon84} proposed a similar origin for
Charon.  If Charon accumulated from a debris disk resulting from such
an impact, the initial eccentricity of Charon's orbit would be
near zero. \citet[][hereafter DPH97]{Dobrovolskis97} were thereby
motivated to
determine the tidal evolution of Charon in a circular orbit to the
current dual synchronous state in a time short compared to the age of
the Solar System (see also \citealt{Farinella79}) as the only possible
outcome of the dissipative process. In a circular orbit,
Charon would reach synchronous rotation very quickly (e.g., DPH97),
and this has generally been assumed
(e.g., \citealt{Peale99}).  However, smoothed particle hydrodynamic
(SPH) simulations by \cite{Canup05} showed
that the results of a nearly intact capture in a glancing encounter
surround the 
$(q,L)$ region of the system much more completely than those of
disk-forming impacts. Therefore, capture where Charon comes off nearly
intact after a glancing impact is favored and non-zero
eccentricity would be more probable. 

We are not aware of any previous attempts to examine the tidal
evolution of Charon's orbit incorporating finite eccentricity.
As we shall see, Charon in an initially eccentric orbit avoids the
almost immediate synchronous rotation heretofore assumed, and the
varied and interesting evolutionary sequences that were suppressed in
the circular orbit evolution are revealed.
Depending on the ratios of rigidity $\mu$ and tidal dissipation
function $Q$ between Pluto and Charon, the eccentricity of Charon's
orbit may either grow or decay during most of the evolution
\citep{Ward06}.
Permanent quadrupole moments of the bodies may also lead to spin-orbit
resonance, and such resonances can have a significant effect on the
orbital evolution.  
 
In the following we tidally evolve the Pluto-Charon system with two
tidal models distinguished by the dependence of the dissipation
function $Q$ on frequency $f$: $Q\propto 1/f$ and $Q=$ constant.
The tidal model developed in Section~\ref{Deltatmodel} has the tidal
distortion of a body responding to the perturbing body a short time
$\Delta t$ in the past.
Constant $\Delta t$ leads to  $Q\propto 1/f$, so we call the $Q\propto
1/f$ model the constant $\Delta t$ model.
In Section~\ref{Qmodel} we develop the equations of evolution for the
constant $Q$ model.
Although neither of these frequency dependences represent the
behavior of real solid materials (e.g., \citealt{Castillo11}) and
although the evolutionary tracks are model dependent,
most if not all of the possible routes from probable initial
configurations to the current equilibrium state are demonstrated.
In Section~\ref{J2C22} we develop the contributions of rotational
flattening $J_2$ and permanent quadrupole moment $C_{22}$ to the
equations of motion.
We describe the adopted system parameters and initial conditions in
Section~\ref{ICs} and the numerical methods in Section~\ref{numerics}.
The results from both the constant $\Delta t$ and constant $Q$ models
with zero $J_{2{\s P}}$ for Pluto and zero $C_{22}$ for both bodies
are shown in
Section~\ref{results1}, and the effects of non-zero $J_{2P}$ and
$C_{22}$ in Section~\ref{results2}, respectively.
The results are discussed in Section~\ref{discussion}, and the
conclusions are summarized in Section~\ref{conclusions}.

\section{TIDAL MODELS}

Tides are raised on Pluto and Charon by each other.
Friction delays the response of the tidal bulge to the tide raising
potential and causes tidal lag.
The lagged bulge leads to angular momentum exchange between
itself and the tide raising body, which leads to rotational and
orbital evolution.

\subsection{Constant $\Delta t$ Tidal Model}
\label{Deltatmodel}

The idea of approximating tidal evolution with a single bulge that
lags by a constant $\Delta t$ was introduced by \citet{Gerstenkorn55},
and developed and used by \citet{Singer68}, \citet{Alexander73},
\cite{Mignard79,Mignard80,Mignard81b}, \cite{Hut81}, and
\cite{Peale05,Peale07}.
The advantage of assuming a single, lagged bulge is that the tidal
forces and torques can be calculated in closed form for arbitrary
eccentricity and inclination.  Either instantaneous or orbit-averaged
tidal forces and torques can be used to determine the evolution.

\begin{figure}[t]
\epsscale{0.7}
\plotone{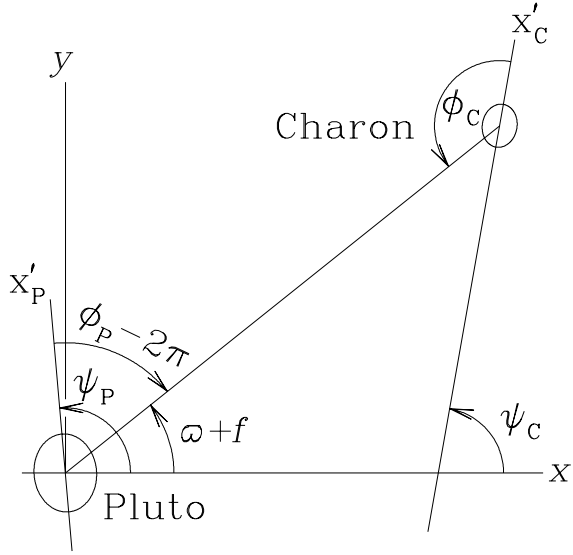}
\caption{Geometry of the Pluto-Charon system with orbit and equator
planes being coplanar. $\psi_i$ are the angles between the axes of
minimum moment of inertia and the inertial $x$ axis, and the $\phi_i$
are the azimuthal angles locating respectively $M_{\s P}$ and $M_{\s C}$
in the other's $x^{\p}$-$y^{\p}$ plane measured counterclockwise from the
$x_i^{\p}$ axes of minimum moment of inertia.\label{fig:geometry}}
\end{figure}

The geometry is illustrated in Fig. \ref{fig:geometry}, where $\psi_{\s P}$
and $\psi_{\s C}$ are the angular displacements of the axes of minimum
moment of inertia from the inertial $x$ axis for Pluto and Charon,
respectively, $\varpi$ is the longitude of periapse, $f$ is the true
anomaly, and $\phi_{\s P}$ and $\phi_{\s C}$ are the azimuthal spherical
coordinates appearing in the potentials for Pluto and Charon,
respectively.
The $x$ and $y$ coordinates are those of Charon relative to Pluto with
the $x$-$y$ plane being the Pluto-Charon orbit plane.
Both spin axes are assumed to be perpendicular to the orbit plane (see
Section \ref{ICs}).
The motion is thereby two dimensional, and the $z$ coordinate is
ignorable.

The tidal contributions to the equations of motion for Charon for this
model are found from the gradient of the tidal potential expanded to
first order in $\Delta t$ \citep{Mignard80, Peale07}:
\begin{eqnarray}
M_{\s PC}\ddot x
&=& -{3 k_{2{\s P}}GM_{\s C}^2R_{\s P}^5 \over r^8} \left[x +
\frac{2{\bf r} \cdot \dot{\bf r} x\Delta t_{\s P}}{r^2} +
(\dot\psi_{\s P} y + \dot x) \Delta t_{\s P} \right] \nonumber\\
& & -{3 k_{2{\s C}}GM_{\s P}^2R_{\s C}^5 \over r^8} \left[x +
\frac{2{\bf r} \cdot \dot{\bf r} x\Delta t_{\s C}}{r^2} +
(\dot\psi_{\s C} y + \dot x) \Delta t_{\s C}\right], \nonumber\\
M_{\s PC}\ddot y
&=& -{3 k_{2{\s P}}GM_{\s C}^2R_{\s P}^5 \over r^8} \left[y +
\frac{2{\bf r} \cdot \dot{\bf r} y\Delta t_{\s P}}{r^2} +
(-\dot\psi_{\s P} x + \dot y) \Delta t_{\s P}\right] \nonumber\\
& & -{3 k_{2{\s C}}GM_{\s P}^2R_{\s C}^5 \over r^8} \left[y +
\frac{2{\bf r} \cdot \dot{\bf r} y\Delta t_{\s C}}{r^2} +
(-\dot\psi_{\s C} x + \dot y) \Delta t_{\s C}\right], \label{eq:gradvt}
\end{eqnarray}
where $G$ is the gravitational constant, ${\bf r}$ and
$\dot{\bf r}$ are the position and velocity of Charon
relative to Pluto, $M_i$, $R_i$, $\dot{\psi}_i$, and $k_{2i}$ are the
mass, radius, spin angular velocity, and second order potential Love
number, respectively, of body $i$ ($= P$ for Pluto and $= C$ for
Charon), and $M_{\s PC}=M_{\s P}M_{\s C}/(M_{\s P}+M_{\s C})$ is the
reduced mass.
The first term on the right hand side of the first (second) equation
in Eq. (\ref{eq:gradvt}) is the $x$-component ($y$-component) of the
force due to the tides raised on Pluto by Charon, and the second
term is the force due to the tides raised on Charon by Pluto.
The equations of motion for the spins are found from the negative of
the torques on the bodies determined from the tidal forces:
\begin{eqnarray}
\mathcal{C}_{\s P}\ddot\psi_{\s P}
&=& -{3k_{2{\s P}}GM_{\s C}^2R_{\s P}^5\Delta t_{\s P} \over r^6}
\left[\dot\psi_{\s P}
  + \frac{-\dot yx + y\dot x}{r^2}\right], \nonumber\\
\mathcal{C}_{\s C}\ddot\psi_{\s C}
&=& -{3k_{2{\s C}}GM_{\s P}^2R_{\s C}^5\Delta t_{\s C} \over r^6}
\left[\dot\psi_{\s C}
  + \frac{-\dot yx + y\dot x}{r^2}\right] ,
\label{eq:tidetorque}
\end{eqnarray}
where $\mathcal{C}_i$ is the moment of inertia of body $i$ about its spin axis.

Eqs. (\ref{eq:gradvt}) and (\ref{eq:tidetorque}) can be used directly
in numerical integration of the equations of motion in Cartesian
coordinates.
Alternatively, one can average the tidal forces and torques over an
orbit to obtain the orbit-averaged equations for the variation of the
spin rate $\dot{\psi}_i$, orbital semimajor axis $a$, and eccentricity
$e$ \citep{Mignard80,Mignard81b}:
\begin{eqnarray}
\frac{1}{n} \left\langle \frac{d \dot{\psi}_i}{dt} \right\rangle &=& -\frac{3G}{C_i a^6} k_{2i} \Delta t_i M_j^2 R_i^5
\left\lbrack f_1(e) \frac{\dot{\psi}_i}{n} - f_2(e) \right\rbrack , \label{fdotpsi} \\
\frac{1}{a} \left\langle \frac{da}{dt} \right\rangle &=&
\frac{6G}{M_{\s PC} a^8} k_{2{\s P}} \Delta t_{\s P} M_{\s C}^2 R_{\s P}^5
\left\lbrack f_2(e) \left( \frac{\dot{\psi}_{\s P}}{n} + A_{\Delta t}
    \frac{\dot{\psi}_{\s C}}{n} \right) - f_3(e) (1+A_{\Delta t}) \right\rbrack , \label{fadot} \\
\frac{1}{e} \left\langle  \frac{de}{dt} \right\rangle &=& \frac{27G}{M_{PC} a^8} k_{2P} \Delta t_{\s P} M_{\s C}^2 R_{\s P}^5
\left \lbrack f_4(e) \frac{11}{18} \left(\frac{\dot{\psi}_{\s P}}{n} +
A_{\Delta t}\frac{\dot{\psi}_{\s C}}{n} \right) - f_5(e) (1+A_{\Delta t})
\right\rbrack , \label{fedot}
\end{eqnarray}
where the subscript $j=$ Charon if $i=$ Pluto, and vice versa,
$\langle \ \rangle$ denotes averaging over an orbit, $n =
[G(M_{\s P}+M_{\s C})/a^3]^{1/2}$ is the mean motion, and
\begin{eqnarray}
f_1(e) &=& \left( 1+ 3e^2 + \frac{3}{8} e^4 \right) \left/ (1-e^2)^{9/2} \right. , \nonumber \\
f_2(e) &=& \left( 1+ \frac{15}{2} e^2 + \frac{45}{8} e^4 + \frac{5}{16} e^6 \right) \left/ (1-e^2)^6 \right. , \nonumber\\
f_3(e) &=& \left( 1+ \frac{31}{2} e^2 + \frac{255}{8} e^4 + \frac{185}{16} e^6 + \frac{25}{64} e^8 \right) \left/ (1-e^2)^{15/2} \right. , \label{fis} \\
f_4(e) &=& \left( 1+ \frac{3}{2} e^2 + \frac{1}{8} e^4 \right) \left/ (1-e^2)^5 \right. , \nonumber \\
f_5(e) &=& \left( 1+ \frac{15}{4} e^2 + \frac{15}{8} e^4 + \frac{5}{64} e^6 \right) \left/ (1-e^2)^{13/2} \right. . \nonumber
\end{eqnarray}
In Eqs. (\ref{fadot}) and (\ref{fedot}),
\begin{equation}
A_{\Delta t} = \frac{k_{2C}}{k_{2P}} \frac{\Delta t_{\s C}}{\Delta t_{\s P}}
\left( \frac{M_{\s P}}{M_{\s C}} \right)^2 \left( \frac{R_{\s C}}{R_{\s P}} \right)^5
\label{eq:ADeltat} 
\end{equation}
is a measure of the relative rate of tidal dissipation in Charon and
Pluto, and the same as $A$ defined in \citet{Mignard80} and in
Eq. (13) of \citet{Touma98}.
$A$ depends on the tidal model and we add subscripts to distinguish them
whenever necessary.

The Love number $k_2$ measures the elastic distortion of the body in
response to the second order spherical harmonic of the deforming
potential.
It can be modeled as (Eq.~[5.6.2] of \citealt{Munk60}; Eq.~[40a] of
\citealt{Peale73})
\begin{equation}
k_2=\frac{k_f}{1+\tilde{\mu}},
\label{lovenumber}
\end{equation}
where $k_f$ is the fluid Love number and $\tilde{\mu}$ is the
effective rigidity. 
The fluid Love number $k_f=3/2$ for homogeneous sphere and its
reduction due to differentiation is sometimes ignored (e.g., DPH97).
The effective rigidity $\tilde{\mu}$ is a dimensionless quantity, and
\begin{equation}
\tilde{\mu}=\frac{19\mu}{2\rho g R}
\end{equation}
for an incompressible homogeneous sphere of radius $R$, rigidity
$\mu$, density $\rho$, and surface gravity $g$ (Eq.~[5.6.1] of
\citealt{Munk60}; Eq.~[4.77] of \citealt{Murray99}).
For small solid body, $\tilde{\mu} \gg 1$ and the approximation
\begin{equation}
k_2\approx \frac{3\rho g R}{19 \mu}
\end{equation}
is commonly used (e.g., \citealp{Peale78, Yoder81, Peale99, Peale07}).
Then
\begin{equation}
A_{\Delta t} \approx \frac{\mu_{\s P}}{\mu_{\s C}} \frac{\Delta t_{\s
C}}{\Delta t_{\s P}} \frac{R_{\s C}}{R_{\s P}} . \label{adtapprox}
\end{equation}

For non-zero eccentricity, the orbit-averaged tidal torque vanishes
at a value of $\dot{\psi}_i > n$, and $\dot{\psi}_i$ goes to an
asymptotic spin rate that increases with eccentricity. This asymptotic
spin is often called a ``pseudo-synchronous'' state.  
The case for Mercury was illustrated by the curve $\delta\sim$
frequency in Fig.~1 of \citet{Goldreich66b}.
Pseudo-synchronous spin rate can be obtained by setting $\langle d
\dot{\psi}_i/dt \rangle = 0$ in Eq. (\ref{fdotpsi}):
\begin{equation}
\frac{\dot{\psi}_{\rm ps}}{n} = \frac{f_2(e)}{f_1(e)}
= 1+ 6e^2 + \frac{3}{8} e^4 + \frac{173}{8} e^6 + O(e^8) .
\label{eq:psidotps}
\end{equation}

For the orbit-averaged equations, we ignore the periapse motion due to
the tidal bulges, as it will be small initially compared to that
caused by the rotational distortion of Pluto (see Section~\ref{ICs}).
Apsidal motion does not affect our discussion as far as the tidal
evolution of Pluto-Charon is concerned, but it does play an important
role in the study of the hypothesis that the small satellites, Nix,
Hydra, Keberos, and Styx, were brought to their current orbits by
mean-motion resonances with Charon (Cheng, Lee and Peale, in
preparation; hereafter paper II).

\subsection{Constant $Q$ Tidal Model}
\label{Qmodel}

A number of authors (e.g.,
\citealt{Goldreich66a,Goldreich66b,Yoder81}) have developed and
applied a tidal model with a constant $Q$, based on the work of
\citet{Kaula64}.
This approach has been used by the previous studies of Pluto-Charon
(\citealp{Farinella79}; DPH97; \citealt{Ward06}).
To develop the equations of evolution for the constant $Q$ model,
expansions in the orbital elements are necessary, and the
orbit-averaged effect on the orbit can be derived from the Gauss
planetary equations.
The truncation of the expansion means the equations are no longer
exact and angular momentum is no longer strictly conserved.

The tidal potential acting on one of the bodies can be written as a
sum of periodic terms.
The frequency of the $(l,m)$ Fourier component of the tidal potential
is
\begin{equation}
\sigma_{lm} = ln - m\dot{\psi},
\end{equation}
where $n$ is the mean motion and $\dot{\psi}$ is the spin angular
velocity of the body.
A phase lag is inserted into the response to each of the periodic
terms in the expansion.
\citet{Zahn77} demonstrated the procedure to expand the lagged tidal
potential in $e$ for the second order spherical harmonic.
The tidal evolution equations in the constant $Q$ model can be derived
from Eqs. (3.6)--(3.8) of \cite{Zahn77}, using the substitution
\begin{equation}
\varepsilon_2^{lm} =
\frac{k_2}{Q} \mbox{sgn} \left( \sigma_{lm} \right) , \label{qes}
\end{equation}
where the tidal coefficient $\varepsilon_2^{lm} = k_2 \sin\alpha$ for
equilibrium tide\footnote{\cite{Zahn77} used the notation $k_2$ for
the apsidal motion constant, which is smaller than the Love number by
a factor 2.},
and the phase angle $\alpha = \mbox{sgn}(\sigma_{lm})/Q$ is assumed to
be small and independent of frequency (except for the sign).
They are
\begin{eqnarray}
\left\langle \frac{ d \dot{\psi}_i}{dt} \right\rangle 
&=& -{3G M_j^2 \over 2\mathcal{C}_i} \frac{k_{2i}}{Q_i}
\frac{R_i^5}{a^6} \left\lbrack \mbox{sgn}(\dot{\psi}_i-n)+e^2D_i + O(e^4)\right\rbrack , \label{qdotpsi} \\
\frac{1}{a}\left\langle \frac{da}{dt} \right\rangle &=& 3n
\frac{k_{2P}}{Q_{\s P}} \frac{M_{\s C}}{M_{\s P}} \left( \frac{R_{\s P}}{a} \right)^5
\left\lbrack \mbox{sgn}(\dot{\psi}_{\s P}-n) + A_Q \mbox{sgn}(\dot{\psi}_{\s C}-n) \right. \nonumber \\
& & \hspace{1.7in} \left. + \: e^2 \left( E_{\s P} + A_QE_{\s C} \right) + O(e^4) \right\rbrack , \label{qadot} \\
\frac{1}{e}\left\langle \frac{de}{dt} \right\rangle &=& n \frac{k_{2P}}{Q_{\s P}} \frac{M_{\s C}}{M_{\s P}} \left( \frac{R_{\s P}}{a} \right)^5 \left\lbrack F_{\s P}+A_QF_{\s C} + O(e^2) \right\rbrack \label{qedot}
\end{eqnarray}
where
\begin{eqnarray}
A_Q &=& \frac{k_{2C}}{k_{2P}} \frac{Q_{\s P}}{Q_{\s C}} \left( \frac{M_{\s P}}{M_{\s C}} \right)^2 \left( \frac{R_{\s C}}{R_{\s P}} \right)^5 \label{eq:AQ}\\
&\approx& \frac{\mu_{\s P}}{\mu_{\s C}} \frac{Q_{\s P}}{Q_{\s C}} \frac{R_{\s C}}{R_{\s P}} . \label{aqapprox}
\end{eqnarray}
Our $A_Q$ is the same as $D$ in Eq. (5) of \cite{Yoder81} and agrees
with the definition of $A$ in \cite{Ward06}.
The above equations conserve the total angular momentum of the system
with an error of $O(e^4)$.

\begin{deluxetable}{crrr}
\tablecolumns{4}
\tablewidth{0pt}
\tablecaption{Coefficients in evolution equations of the constant $Q$ model
\label{coeffQ}}
\tablehead{
\colhead{$\dot{\psi}_i/n$} & \colhead{$D_i$} & \colhead{$E_i$} & \colhead{$F_i$}
}
\startdata
$>3/2$ & $15/2$ & $51/4$ & $57/8$ \\
$=3/2$ & $-19/4$ & $-45/8$ & $-33/16$ \\
$>1$ and $<3/2$ & $-17$ & $-24$ & $-45/4$ \\
$=1$ & $-12$ & $-19$ & $-21/2$ \\
$>1/2$ and $<1$ & $-7$ & $-14$ & $-39/4$ \\
\enddata
\end{deluxetable}

The coefficients $D_i$, $E_i$, and $F_i$ depend on the spin of $M_i$,
as listed in Table~\ref{coeffQ}.
Discontinuous dependence on $\dot{\psi}_i/n$ of these coefficients
arises from the sign changes of $\sigma_{lm}$.
The coefficients for $\dot{\psi}_i/n>3/2$ and $=1$ have been
documented in the literature:
$D_i$ in \cite{Ferraz-Mello08} and $E_i$ and $F_i$ in \cite{Peale80} and
\cite{Yoder81}.
$E_i = -19$ for synchronous rotation differs from the value in the
literature, which included the effect of permanent quadrupole moment
(see Section 12.1 of \citealt{Ferraz-Mello08} and footnote 6 of
\citealt{Efroimsky09}).
We treat permanent quadrupole moment separately in the next
subsection.

A closer inspection of Eq.~(\ref{qdotpsi}) reveals that the asymptotic
spin rate ($\langle d\dot{\psi}_i/dt\rangle = 0$) of the body
discontinuously depends on the orbital eccentricity.
Eq.~(\ref{qdotpsi}) changes sign when $e$ increases from below
$1/\sqrt{17} = 0.243$ to above.
A body in asymptotic spin would then increase its spin from
synchronous to $3n/2$ in the spin evolution timescale.
The discontinuity occurs at $e=0.235$ if we take higher order terms in
$e$ into consideration \citep{Goldreich66b}.
Fig.~1 of \citet{Goldreich66b} showed the next discontinuity as well.
We calculate its position to be at $e\approx0.36$, using coefficients
up to $O(e^4)$ in Eq.~(80) of \cite{Efroimsky09}.
As the coefficients of Eq.~(\ref{qadot}) and (\ref{qedot}) for higher
order terms in $e$ are not easily derivable, we restrict our analysis
to the current order and note that our results from this model are
qualitatively inaccurate for $e \gtrsim 0.36$.

Before we turn to the effects of rotation induced oblateness and
permanent axial asymmetry in the next subsection, we note that the
equations of \cite{Zahn77} can also be used to derive the evolution
equations expanded in eccentricity for the constant $\Delta t$ model.
For small phase lag, if we let $\alpha = \Delta t \sigma_{lm}$ and
$\varepsilon_2^{lm} = k_2 \Delta t \sigma_{lm}$, then
Eqs. (3.6)--(3.8) of \citet{Zahn77} give
\begin{eqnarray}
\frac{1}{n} \left\langle \frac{d \dot{\psi}_i}{dt} \right\rangle &=& -\frac{3G}{\mathcal{C}_i a^6} k_{2i} \Delta t_i M_j^2 R_i^5
\left\lbrack  \left( 1+ \frac{15}{2} e^2 \right) \frac{\dot{\psi}_i}{n} - \left(  1+\frac{27}{2}e^2 \right) + O(e^4) \right\rbrack , \label{tdotpsi} \\
\frac{1}{a} \left\langle \frac{da}{dt} \right\rangle &=& \frac{6G}{M_{PC} a^8} k_{2P} \Delta t_{\s P} M_{\s C}^2 R_{\s P}^5 
\left\lbrack \left( 1+\frac{27}{2}e^2 \right) \left( \frac{\dot{\psi}_{\s P}}{n} + A_{\Delta t} \frac{\dot{\psi}_{\s C}}{n} \right) \right. \nonumber \\
& & \hspace{1.7in} \left. - \left(1+ 23e^2\right) \left(1+A_{\Delta t} \right)+ O(e^4) \right\rbrack , \label{tadot} \\
\frac{1}{e} \left\langle  \frac{de}{dt} \right\rangle &=& \frac{27G}{M_{PC} a^8} k_{2P} \Delta t_{\s P} M_{\s C}^2 R_{\s P}^5
\left \lbrack \frac{11}{18} \left( \frac{\dot{\psi}_{\s P}}{n} + A_{\Delta t} \frac{\dot{\psi}_{\s C}}{n} \right) - \left(1+A_{\Delta t} \right) + O(e^2) \right\rbrack . \label{tedot}
\end{eqnarray}
These equations agree with the exact equations
(Eqs.~[\ref{fdotpsi}]--[\ref{fedot}]) to $O(e^2)$, as expected.
We will compare the results from these equations and from the exact
equations to get an idea how good the results from the
$O\left(e^2\right)$ equations of the constant $Q$ model are for $e \la
0.36$.

\subsection{Rotational Flattening and Permanent Quadrupole Moment}
\label{J2C22}

Rotational flattening and internal uneven mass distribution give
non-zero gravitational harmonic coefficients $J_2 =
[\mathcal{C} - (\mathcal{A} + \mathcal{B})/2]/(M R^2)$ and
$C_{22} =
(\mathcal{B} - \mathcal{A})/(4 M R^2)$, where $\mathcal{A} \le
\mathcal{B} \le \mathcal{C}$ are the principal moments of inertia.
The contributions of these terms to the equations of motion of a
spinning rigid body $i$ orbiting another body $j$ were derived by
\cite{Touma94b} and in Chapter 5 of \cite{Murray99}, which include the
change in the spin rate of the body and the feedback on the orbit.
In our aligned configuration with the rigid body $i$ rotating about
its axis of maximum moment of inertia, the motion of the rigid body is
confined on a plane and both equations can be greatly simplified.
The spin equations become
\begin{equation}
\mathcal{C}_i \ddot{\psi}_i = -6\frac{G M_j}{r^5} C_{22i} M_i R_i^2
\left[(x^2-y^2)\sin{2\psi_i}-2xy\cos{2\psi_i}\right], \label{psiddot}
\end{equation}
where the notation is as shown in Fig.~\ref{fig:geometry}.
The contributions of $J_2$ and $C_{22}$ to the acceleration of Charon
relative to Pluto are
\begin{eqnarray}
M_{\s PC}\ddot x&=&
GM_{\s P}M_{\s C}\Bigg{\{}-\frac{3J_{2{\s P}}R_{\s P}^2x}{2r^5}
 \nonumber\\
&+&3(C_{22{\s P}}R_{\s P}^2\cos{2\psi_{\s P}}+C_{22{\s C}}R_{\s
C}^2\cos{2\psi_{\s
C}})\left[\frac{2}{r^5}-\frac{5(x^2-y^2)}{r^7}\right] x \nonumber\\
&+&3(C_{22{\s P}}R_{\s P}^2\sin{2\psi_{\s P}}+C_{22{\s C}}R_{\s
C}^2\sin{2\psi_{\s C}})\left[\frac{2}{r^5}-
\frac{10x^2}{r^7}\right] y \Bigg{\}}\nonumber\\ 
M_{\s PC}\ddot y&=&
GM_{\s P}M_{\s C}\Bigg{\{}-\frac{3J_{2{\s P}}R_{\s P}^2y}{2r^5}
 \nonumber\\
&+&3(C_{22{\s P}}R_{\s P}^2\cos{2\psi_{\s P}}+C_{22{\s C}}R_{\s
C}^2\cos{2\psi_{\s
C}})\left[\frac{-2}{r^5}-\frac{5(x^2-y^2)}{r^7}\right] y \nonumber\\
&+&3(C_{22{\s P}}R_{\s P}^2\sin{2\psi_{\s P}}+C_{22{\s C}}R_{\s
C}^2\sin{2\psi_{\s C}})\left[\frac{2}{r^5}-
\frac{10y^2}{r^7}\right] x \Bigg{\}},
\label{eq:motioneq}
\end{eqnarray}
where $J_{2{\s C}}$ of Charon is omitted.

\section{PARAMETERS AND INITIAL CONDITIONS}
\label{ICs}

In all our calculations, Charon is assumed to start in an eccentric
orbit with semimajor axis $a = 4R_{\s P}$ (Pluto radii), consistent
with its origin in a nearly intact capture from a glancing impact on
Pluto \citep{Canup05}.
Since a significant portion of the angular momentum of the impactor is
transferred to the spin of the target in the collision (see Table 1 of
\citealt{Canup05}), the spin axis of Pluto should be close to being
perpendicular to Charon's orbit initially.
The spin axis of Charon, which could be inclined from the orbit normal
initially, would quickly approach a Cassini state with the spin axis
close to the orbit normal, on a timescale comparable to the timescale
for Charon to reach asymptotic spin rate in at least one of the tidal
models (see Eq.~[53] of \citealt{Hut81}).
Moreover, tidal evolution of the orbit and spin rates is unaffected to
first order in orbital inclination
\citep{Hut81,Ferraz-Mello08,Efroimsky09}.
Thus, to reduce the complexity and the parameter space of the problem,
we only examine the aligned configuration of Pluto-Charon, where the
orbit normal aligns with their spin axes.

We specify $a$, $e$, and $\dot{\psi}_{\s C}$ as our initial conditions, and
initial $\dot{\psi}_{\s P}$ is calculated by assuming the same total
angular momentum as the current Pluto-Charon system:
\begin{equation}
L = \mathcal{C}_{\s P} \dot{\psi}_{\s P} + \mathcal{C}_{\s C} \dot{\psi}_{\s C} +
    M_{PC} n a^2 \sqrt{1-e^2}
= \left(\mathcal{C}_{\s P} + \mathcal{C}_{\s C} + M_{PC} a_0^2\right)n_0 ,
\label{angmot}
\end{equation}
where $a_0$ and $n_0$ are the current separation and mean motion of
Pluto-Charon, respectively.
Charon's spin angular momentum is always small compared to the total.
The numerical value
$L = 6.00 \times 10^{37}\,{\rm g}\,{\rm cm}^2\,{\rm s}^{-1}$
is determined under the assumption that the dimensionless moments of
inertia $\bar{\mathcal{C}}_{\s P} =
\mathcal{C}_{\s P}/(M_{\s P} R_{\s P}^2) = 0.328$ and
$\bar{\mathcal{C}}_{\s C} = \mathcal{C}_{\s C}/(M_{\s C} R_{\s C}^2) = 0.4$.
The numerical value for $\bar\mathcal{C}_{\s P}$ follows from a
two-layer model (DPH97) with a rocky core (density
$3.0\,{\rm g}\,{\rm cm}^{-3}$) and an icy mantle (density
$1.0\,{\rm g}\,{\rm cm}^{-3}$). Pluto should be so
differentiated by the giant impact if it was not earlier
\citep{McKinnon89, Canup05}. 
The internal structure of Charon is uncertain (e.g.,~\citealp[Section
6.2]{McKinnon08}) and we assume that Charon remains homogeneous, i.e.,
$\bar{\mathcal{C}}_{\s C}=0.4$ (\citealp{McKinnon89}; DPH97).

Pluto's Love number $k_{2P}=0.058$, computed using
Eq.\ (\ref{lovenumber}) with $k_{fP} = 3/2$ and $\mu_{\s P} = 4\times10^{10}
\,{\rm dynes}\,{\rm cm}^{-2}$ of water ice.
For the constant $\Delta t$ model, we adopt $\Delta t_{\s P} = 600$
seconds, same as that for the Earth \citep{Mignard80, Touma94a}.
For the constant $Q$ model, we adopt $Q_{\s P} = 100$, as typically assumed
for solid bodies (DPH97; Tables 4.1 and 4.2 of \citealt{Murray99}).
These parameters are fixed throughout the tidal evolution.
Our incomplete knowledge of the physics of tides and of the
composition and internal structure of Pluto means that the actual
values of these parameters are not well constrained.
Yet uncertainties in these parameters only affect the overall
timescale of tidal evolution, as far as spin-orbit resonance is not
included.
If the rigidities ($\mu$) and dissipation ($\Delta t$ or $Q$) of Pluto
and Charon are comparable, from Eqs.~(\ref{adtapprox}) and
(\ref{aqapprox}), one would expect $A_{\Delta t}$ and $A_Q \approx
R_{\s C}/R_{\s P} \approx 1/2$.
Alternatively, if the Love numbers ($k_2$) and dissipation ($\Delta t$
or $Q$) of Pluto and Charon are comparable, one would expect
$A_{\Delta t}$ and $A_Q \approx (M_{\s P}/M_{\s C})^2 (R_{\s C}/R_{\s P})^5
\approx 3$.
In our integrations, we focus on those values of $A_{\Delta t}$ and
$A_Q$ that can keep $e$ roughly constant until the end of the tidal
evolution.
If $A$ is too large, the orbit circularizes quickly, and the tidal
evolution would be similar to that already studied by DPH97.
If $A$ is too small, $e$ can approach $1$, and the system can become
unstable (see Section~\ref{results}).
The evolutions with $e$ roughly constant throughout most of the tidal
evolution are also the most likely ones that allow migration of the
small satellites in resonances, since resonances cannot be maintained
if $e$ is too small and become unstable if $e$ is too large
\citep{Ward06,Lithwick08a}.
We discuss the details in paper II.

We estimate the largest value that $J_{2{\s P}}$ of Pluto is likely to
be by the hydrostatic value just after the impact that captured
Charon.
For rotation about the axis of maximum moment of inertia, the changes
in the principal components of the inertia tensor from rotation are
given by (e.g., \citealt{Peale73})
\begin{eqnarray}
\Delta\mathcal{A} = \Delta\mathcal{B}
&=& - {k_{fP} R_{\s P}^5 {\dot\psi}_{\s P}^2\over 9 G} ,
\nonumber \\
\Delta\mathcal{C} &=& + {2 k_{fP} R_{\s P}^5 {\dot\psi}_{\s P}^2\over 9 G} ,
\label{eq:inertiatensor}
\end{eqnarray}
where $k_{fP}$ is the fluid Love number of Pluto.
Then
\begin{equation}
J_{2{\s P}}
= \frac{\Delta \mathcal{C} - (\Delta \mathcal{A} + \Delta\mathcal{B})/2}
  {M_{\s P}R_{\s P}^2}
= \frac{k_{fP} R_{\s P}^3\dot\psi_{\s P}^2}{3 G M_{\s P}} .
\label{eq:j2}
\end{equation}
For an initial Pluto spin period of $3.15$ hours, compatible with our
typical initial conditions of $a = 4R_{\s P}$, $e = 0.2$, and
$\dot{\psi}_{\s C} = 2n$, $J_{2{\s P}} \approx 0.17$--$0.27$ if
$k_{fP} \approx 1$ (by analogy with the Earth) to $3/2$ (for
homogeneous sphere).
The large value means that inclusion of higher order terms in the
rotational distortion would be appropriate, but we have not tried to
obtain a more accurate estimate of $J_{2P}$.
The large $J_{2P}$ reflects the  fact that the estimated spin of Pluto
for $a \sim 4 R_{\s P}$ is close to the limit of rotational instability
according to the ratio of rotational to gravitational binding energy
for a homogeneous Pluto (DPH97 and references therein).

Satellite motion around an oblate body deviates from that of
Keplerian.
If $C_{22P}$ is negligibly small ($\sim 10^{-5}$ or less), corrections
on the mean motion and the rate of periapse precession are given by
Eqs.\ (6.244) and (6.249) of \cite{Murray99}:
\begin{eqnarray}
n^2 &=& \frac{G(M_{\s P}+M_{\s C})}{a^3} \left\lbrack 1 + \frac{3}{2} J_{2P}
\left( \frac{R_{\s P}}{a} \right)^2 + O \left( \frac{R_{\s P}}{a} \right)^4
\right\rbrack , \label{j2meanmot} \\
\dot{\varpi} &=& \left[\frac{G(M_{\s P}+M_{\s C})}{a^3}\right]^{1/2}
\left\lbrack \frac{3}{2} J_{2P} \left( \frac{R_{\s P}}{a} \right)^2 + O
\left( \frac{R_{\s P}}{a} \right)^4 \right\rbrack . \label{j2prec}
\end{eqnarray}
We estimate the correction on the mean motion to be less than $\sim 1\%$
initially.
It decreases further with $\dot{\psi}_{\s P}$ and hence can be ignored.
On the other hand, the precession due to the oblateness of Pluto is
much larger than that from the oblateness of Charon and tidal
deformation during most of the evolution, and the remnant $J_{2{\s
P}}$ supported by internal stress is likely to be greater than the
hydrostatic value and the tidal value for the current configuration.

The value of initial $J_{2P}$ we use in our integrations is $J_{2P} =
0$ or $0.1$, representing the two extreme cases with no or very fast
precession of the orbit.
As Charon moves outward, we assume that $J_{2P}$ decreases with
$\dot{\psi}_{\s P}^2$ (Eq.~[\ref{eq:j2}]).
We ignore the smaller effect of Charon's $J_2$, and we choose $C_{22P}
= C_{22C} = 0$ or $10^{-5}$ for the integrations, where the latter
value is comparable to the measured values of other nearly spherical
solid bodies in the Solar System.

\section{NUMERICAL METHODS}
\label{numerics}

\subsection{Runge-Kutta Codes}

For the calculations without $J_{2P}$ and $C_{22}$, the most efficient
way to evolve Pluto-Charon is to solve the orbit-averaged tidal
evolution equations: Eqs.\ (\ref{fdotpsi})--(\ref{fedot}) and
Eqs.\ (\ref{qdotpsi})--(\ref{qedot}) in the two tidal models.
We use the 4th order Runge-Kutta method with an adaptive time step
(e.g., \citealt{Press92}).
We find that it is simpler and more accurate to treat
$\dot{\psi}_i/n$ as variables rather than $\dot{\psi}_i$ in these
codes, through
\begin{equation}
\left\langle \frac{d}{dt} \left( \frac{\dot{\psi}_i}{n} \right) \right\rangle = \frac{1}{n} \left\langle \frac{d \dot{\psi}_i}{dt} \right\rangle + \frac{3}{2}\frac{\dot{\psi}_i}{n} \frac{1}{a} \left\langle \frac{da}{dt} \right\rangle \label{dpsidotn}.
\end{equation}
Discontinuities of coefficients in the equations of the constant $Q$
model need special treatment to prevent the adaptive time step
algorithm from crashing when the system comes across them.
One can smooth the discontinuities by assuming that $Q$ has a very
weak power dependence on frequency using Eq.\ (86) of \cite{Efroimsky09},
but we find it convenient to use a simple smoothing function
recommended by \citet{Rauch99}.
We modify the smoothing function in their Eq.\ (29) to give a step
from $-1$ (at $\delta=-\epsilon$) to $1$ (at $\delta=\epsilon$) with
controllable steepness:
\begin{equation}
\kappa(\delta,\epsilon) = \tanh\left[
\frac{4\delta/\epsilon}{1-(\delta/\epsilon)^2}\right].
\end{equation}
Here $\delta$ denotes the percentage difference of $\dot{\psi}_i/n$
from discontinuity, and $\epsilon$ is an adjustable parameter within
which smoothing is applied.
The advantage of this smoothing function is that all orders of
derivatives vanish at both the beginning and end of the transition
over the discontinuity.
The position and amplitude of the smoothing are transformed to replace
the discontinuous step, such that when $|\delta |< \epsilon$, the
coefficients would be calculated using the above equation instead.
The parameter $\epsilon$ can be set to at most 20\% without
overlapping for the discontinuities at $\dot{\psi}_i/n= 1$ and 3/2.
We use $\epsilon=1$\% in most cases.
In the $N$-body codes described in the next subsection, smoothing of
the discontinuities in the constant $Q$ model can be turned off.
We compare the results from $N$-body calculations with and without
smoothing and find very small differences for our typical $\epsilon$
of $1\%$.

We perform three classes of tests on the Runge-Kutta codes.
We test the implementation of each equation separately in the first
class of test.
By setting the right hand side of all but one of the tidal evolution
equations to zero, analytical solutions are available for each
equation, except for Eq.~(\ref{fedot}).
In the constant $Q$ model, analytical solution of each equation is
valid only for no discontinuity crossing.
We monitor the angular momentum budget of the system as a second class
of test.
In the constant $\Delta t$ model, total angular momentum of the system
is conserved to better than the tolerance parameter ($\sim 10^{-10}$ for the
results presented in Section~\ref{results}) of the adaptive time step
algorithm throughout the evolution to the current dual synchronous
state.
In the constant $Q$ model, Eqs.~(\ref{qdotpsi})--(\ref{qedot}) do not
conserve the total angular momentum of the system.
The discrepancy becomes worse when eccentricity is large.
Smoothing the discontinuities also contributes to the error.
When tolerance is small (including the adopted $\sim 10^{-10}$), the
angular momentum error of the system is dominated by the order of the
equations in $e$.
The third class of test aims at testing the implementation of
Eq.~(\ref{fedot}).
In the constant $\Delta t$ model, \citet{Hut81} found that the tidal
evolution between a body and a point mass (i.e., tides are raised on
one body only) depends on the value of $\tilde\alpha$ only for given
initial $a$ and $e$, where $\tilde\alpha$ is the ratio of orbit to
spin angular momentum at the dual synchronous state.
Figs.~5--8 of \citet{Hut81} show the flow lines of tidal evolution in
the ($e, a/a_0$) space for four different values of $\tilde\alpha$,
where $a_0$ is the separation in the dual synchronous state.
By treating either Pluto or Charon as a point mass, we are able to
reproduce the flow lines in all four figures by choosing suitable
initial conditions, including those with initial $e > 0.9$.

\subsection{$N$-body Codes}

As the effects of $C_{22}$ are on suborbital timescale, no analytic,
orbit-averaged equations are available for evolving spins. 
Thus, to study the consequences of non-zero quadrupole moments
represented by $J_2$ and $C_{22}$ on the tidal evolution, it is
necessary to perform $N$-body integrations.

For the constant $\Delta t$ model, the equations of motion in
Cartesian coordinates with the instantaneous tidal forces and torques
and the effects of $J_{2P}$ and $C_{22}$ (Eqs.~[\ref{eq:gradvt}],
[\ref{eq:tidetorque}], [\ref{psiddot}], and [\ref{eq:motioneq}]) can
be integrated directly.
We have written a code implementing these equations using the
Bulirsch-Stoer method, and its accuracy is verified by the
conservation of angular momentum to more than 8 significant figures
for integration from typical initial conditions to the dual
synchronous state.
This Bulirsch-Stoer code has the advantage of solving the exact
equations of motion.
However, it is slow for realistic values of $\Delta t$, because the
tidal forces and torques are computed many times over an orbit, even
though they are weak and affect the evolution only on the tidal
evolution timescale.
In addition, this approach does not work for the constant $Q$ model,
where the instantaneous tidal forces and torques are not available.

Thus, for both tidal models, we also modify the Wisdom-Holman
\citeyearpar{Wisdom91} integrator in the 
SWIFT\footnote{See http://www.boulder.swri.edu/$\sim$hal/swift.html.}
package \citep{Levison94} to simulate the tidal, rotational and
axial-asymmetry effects.
The SWIFT package allows non-zero $J_2$ for the central body (i.e.,
Pluto in our case). For non-zero initial $J_{2P}$, we adjust it to
decrease $\propto 
\dot{\psi}_{\s P}^2$ throughout the evolution according to
Eq.~(\ref{eq:j2}).
Other effects are imposed following the approach of \citet{Lee02}:
\begin{equation}
E_{e}(\frac{m\tau}{2}) E_{a}(\frac{m\tau}{2}) E_{\dot{\psi}}(\frac{m\tau}{2})
\underbrace{E_{C_{22}}(\frac{\tau}{2})E_{\rm rot}(\tau)E_{\rm WH}(\tau)E_{C_{22}}(\frac{\tau}{2})}_{m\ {\rm copies}}
E_{\dot{\psi}}(\frac{m\tau}{2}) E_{a}(\frac{m\tau}{2}) E_{e}(\frac{m\tau}{2}).
\end{equation}
Here $E_{\rm WH}(\tau)$ denotes a complete step of time step $\tau$ in
the Wisdom-Holman scheme, and the other $E$'s are evaluations for
each effect.
In each evaluation, only those variables concerned are evolved and
others are kept constant.

We follow the method presented by \citet{Touma94b} for rotation and
the effects of $C_{22}$.
We represent the pointing directions of the long axes of both
bodies by unit vectors.
$E_{\rm rot}(\tau)$ rotates them according to the instantaneous
$\dot{\psi}_i$ of the bodies.
The bodies are treated as axisymmetric in $E_{\rm WH}(\tau)$, and
hence $E_{\rm WH}(\tau)$ commutes with $E_{\rm rot}(\tau)$.
$E_{C_{22}} (\tau/2)$ changes the spins and velocities of the bodies
according to Eqs.~(\ref{psiddot}) and (\ref{eq:motioneq}).
These two evaluations, $E_{\rm rot}$ and $E_{\rm C_{22}}$, are the
spin analog to the leapfrog integration, except that the feedback on
the orbit has to be included.
Our $N$-body simulations typically start with the long axis of both
Pluto and Charon pointing along the inertial $x$-axis and Charon at
periapse on the $x$-axis.

The substeps $E_{a}$, $E_{e}$, and $E_{\dot{\psi}}$ correspond to the
changes in $a$, $e$, and $\dot{\psi}_i$ in the orbit-averaged tidal
evolution equations.
Their sequence is chosen such that the computationally expensive $f_i
(e)$ in Eq.~(\ref{fis}) are calculated once only in each half-step.
Eq.~(\ref{dpsidotn}) is not used here as $E_{a}$ and $E_{\dot{\psi}}$
are applied sequentially.
In the constant $\Delta t$ model, solving Eq.~(\ref{fadot})
analytically either involves complicated expressions or
transformations back and forth between $\dot{\psi}_i$ and
$\dot{\psi}_i/n$ in each step, and we use explicit midpoint method in
$E_{a}$ and $E_{e}$ and analytical solution in $E_{\dot{\psi}}$.
In the constant $Q$ model, analytical expressions for all the substeps
are available, assuming all coefficients are constant during the step.

The parameter $m$ is an integer, and tides should be applied on tidal
evolution timescale by using a large $m$.
This is done for numerical efficiency and to reduce roundoff error.
There is an error introduced by the conversion between the positions
and velocities and the osculating orbital elements.
In the regime of eccentricity and step size of our problem, this error
is tested to be secularly increasing with the number of tidal steps
taken, which can be significantly reduced by the use of a large $m$.

We use an initial $\tau=10^3$ seconds, which is about 60 steps per
orbit for initial $a\approx4R_{\s P}$, and an initial $m\tau=10^5$
seconds.
Since $m\tau$ is kept constant and tidal evolution is proportional to
a large negative power of $a$, the relative angular momentum error
introduced by our second order solution saturates at $\sim 10^{-7}$
soon after $a$ starts to increase.
We integrate the system up to a point when $a$ has increased by a
significant factor ($a \sim 11 R_{\s P}$), then we increase $\tau$ and $m$.
The step size $\tau$ is increased by a factor of 5 to give a similar
number of steps per orbit as initially.
The tidal step size $m\tau$ is increased by a factor of 100, which is
smaller than one would use to keep the right hand side of the tidal
equations comparable in magnitude as initially.
We choose this factor of 100 so that the increase in the step size
does not further increase the already saturated relative angular
momentum error of the system.
Because of the slower evolution rate in the constant $Q$ model, we
increase the step size once more, when $a$ and $e$ are around their
maximum for the smallest $A_Q$ (see below).
This time $\tau$ is increased by a factor of 2, and the tidal step
size is increased by a factor of 10.

We perform several tests on the $N$-body codes.
When $J_{2P}$ and $C_{22}$ are set to zero, results from the
Wisdom-Holman and Bulirsch-Stoer codes coincide with those from the
Runge-Kutta codes. 
For uniform rotation, the pointing directions of unit vectors are
tested to change at the expected rate.
Precession due to an oblate Pluto is tested to agree with the expected
rate given by Eq.~(\ref{j2prec}).
The $C_{22}$ effects without tides are tested to conserve total
angular momentum of the system.
Finally, we compare results from the Wisdom-Holman and Bulirsch-Stoer
codes, and they agree in all cases examined.

\section{RESULTS}
\label{results}

\subsection{Tidal Evolution with Zero $J_{2P}$ and $C_{22}$}
\label{results1}

In this subsection, we present the tidal evolution of Pluto-Charon for
both the constant $\Delta t$ and constant $Q$ models, with $J_{2P} =
0$ and $C_{22} = 0$ for both Pluto and Charon.
The results are obtained using the Runge-Kutta codes, unless otherwise
specified.

\begin{figure}
\epsscale{0.65}
\plotone{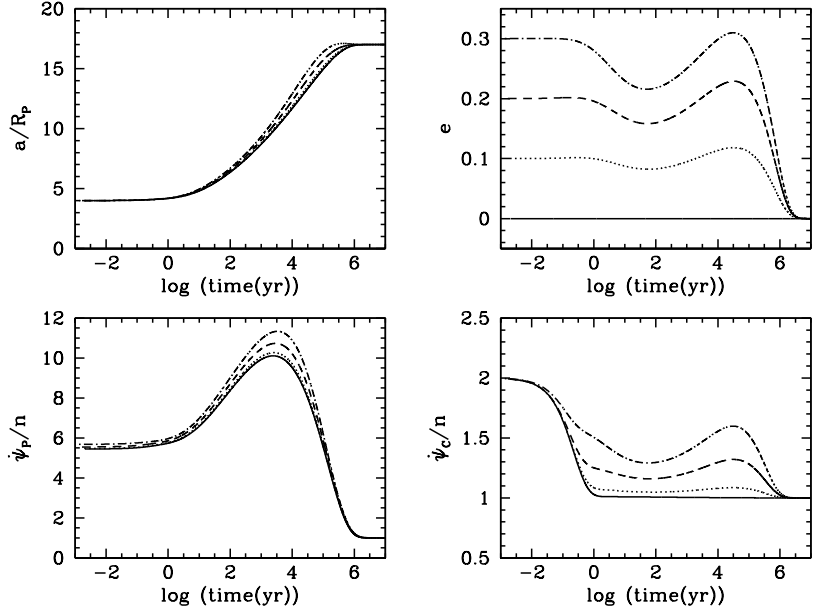}
\caption[Evolution in the constant $\Delta t$ model with a range of initial $e$]
{Evolution in the constant $\Delta t$ model with initial $e$ from 0 to
0.3 (in steps of 0.1) and $A_{\Delta t} = 10$.
The panels show the orbital semimajor axis $a$ in units of Pluto
radius $R_{\s P}$, orbital eccentricity $e$, and the spin angular
velocities of Pluto and Charon, $\dot\psi_{\s P}$ and $\dot\psi_{\s C}$,
in units of the mean motion $n$.}
\label{fe0123}
\end{figure}

\begin{figure}
\epsscale{0.65}
\plotone{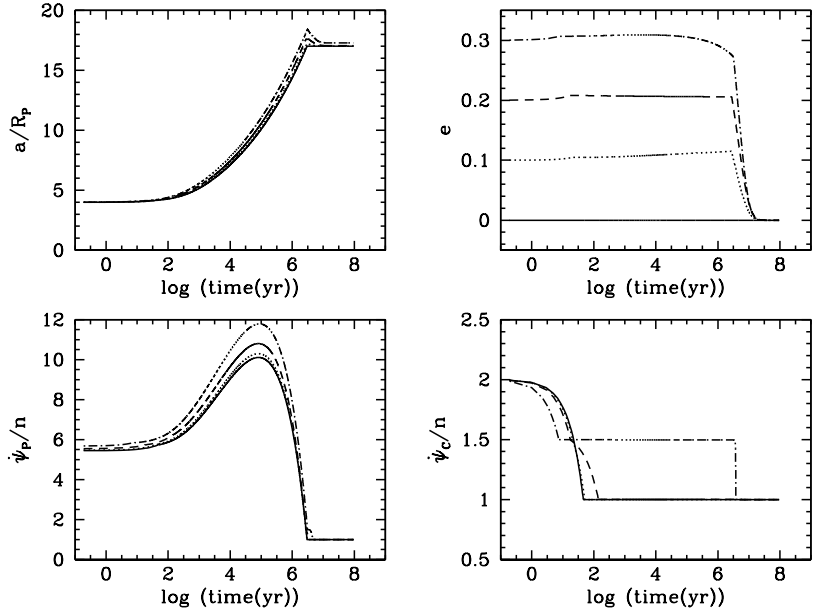}
\caption[Evolution in the constant $Q$ model with a range of initial $e$]{Evolution in the constant $Q$ model with initial $e$ from 0 to 0.3 (in steps of 0.1). $A_Q=1.15$ for initial $e=0.3$ and $A_Q= 0.65$ for other initial $e$.}
\label{qe0123}
\end{figure}

Figs.~\ref{fe0123} and \ref{qe0123} show the evolution in both
tidal models using our typical initial conditions of $a=4R_{\s P}$ and
$\dot{\psi}_{\s C}=2n$ for a range of initial eccentricities.
The relative rate of tidal dissipation in Charon and Pluto,
$A_{\Delta t}$ and $A_Q$ defined in Eqs.~(\ref{eq:ADeltat}) and
(\ref{eq:AQ}), is chosen such that $e$ is kept roughly constant
throughout most of the evolution ($A_{\Delta t} = 10$, and $A_{Q} =
0.65$ and $1.15$ for initial $e \le 0.2$ and $e = 0.3$, respectively).
Note that all evolutions shown reach the current dual synchronous
state of Pluto-Charon, as predicted by DPH97.
For $e=0$, the spin of Charon drops to synchronous quickly, as
estimated by DPH97.
However, the assumption that the spin of Charon is synchronous
throughout most of the evolution does not necessarily hold for
non-zero $e$.
For constant $\Delta t$, the spin of Charon achieves the
pseudo-synchronous state quickly instead, and evolves according to $e$
afterwards (Eq.~[\ref{eq:psidotps}]).
For constant $Q$, the asymptotic spin rate for $e > 0.235$ is no
longer synchronous but $3n/2$, as mentioned in Section~\ref{Qmodel}.
Hence, for larger initial $e$ and spin of Charon above $3n/2$, the
spin of Charon first reaches and stays at $3n/2$, and falls to
synchronous depending on the eccentricity evolution (see, e.g., the
evolution with initial $e = 0.3$ in Fig.~\ref{qe0123}).
The angular momentum carried in the rotation of Charon is small
throughout the evolution, as the moment of inertia of Charon is much
smaller than that of Pluto (a factor of $\sim30$) and its spin stays
within a factor of two of synchronous for mild eccentricity
($e\lesssim0.4$).
Note that $\dot{\psi}_{\s P}/n$ rises to $>10$ (higher for larger
eccentricity) before falling to synchronous, even though the rotation
rate of Pluto is monotonically decreasing.
This initial rise in $\dot{\psi}_{\s P}/n$ is due to $n$ decreasing
faster than $\dot{\psi}_{\s P}$.

The tidal evolution can be drastically affected by the relative rate
of tidal dissipation in Charon and Pluto ($A_{\Delta t}$ or $A_Q$).
If the spin angular velocity of Pluto sufficiently exceeds the orbital
angular velocity of Charon at periapse, the maximum tide at that point
in the orbit gives Charon a kick that tends to increase the
eccentricity.
Otherwise tides raised on Pluto will decrease the eccentricity (see
Eqs. [\ref{fedot}] and [\ref{qedot}]).
Since Charon's rotation stays within a factor of two of synchronous
for mild eccentricity, tides raised on Charon typically damp the
eccentricity.
Since Pluto will be initially spinning very fast, we expect there will
be a tendency for tides raised on Pluto to increase the orbital
eccentricity that will be counteracted by tides raised on Charon
tending to decrease the eccentricity.
Which wins depends on the value of $A$.

\begin{figure}
\epsscale{0.7}
\plotone{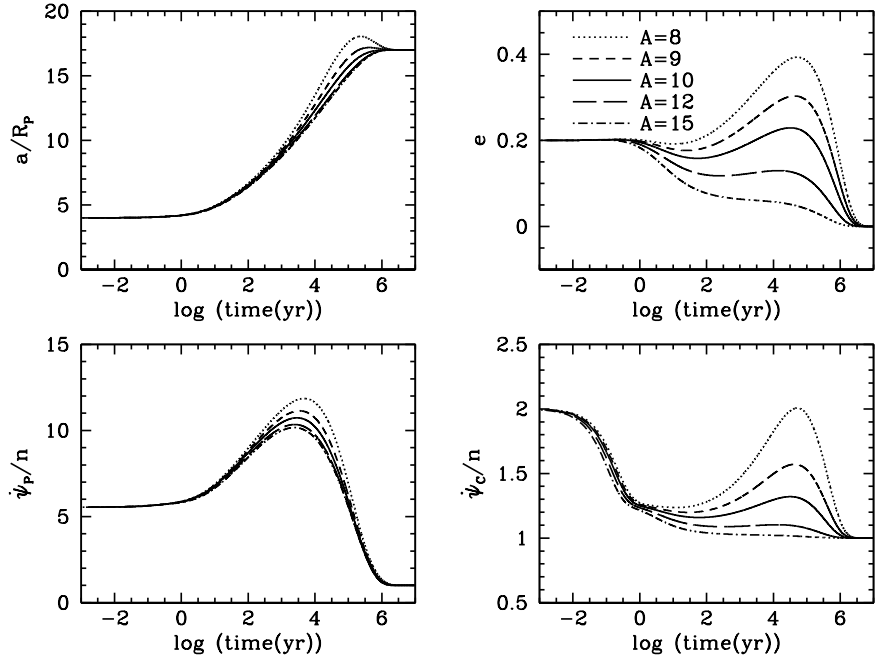}
\caption[Evolution in the constant $\Delta t$ model with initial $e=0.2$]{Evolution in the constant $\Delta t$ model with initial $e=0.2$ and $A_{\Delta t}=8 - 15$.}\label{fe2}
\end{figure}

Fig.~\ref{fe2} shows the evolution in the constant $\Delta t$ model
with initial $e=0.2$ and a range of $A_{\Delta t}$.
While $e$ can be kept more or less constant as $a$ increases if
$A_{\Delta t} = 10$ is used, larger (smaller) $A_{\Delta t}$
would result in $e$ decreasing (increasing) throughout most of the
evolution.
If the eccentricity is still large ($e\gtrsim0.3$) when $a$ reaches
the current value ($17 R_{\s P}$), then by the
conservation of angular momentum, it is expected that $a$ would
overshoot before coming back to the current value when $e$ decays.
This is seen clearly in the case with $A_{\Delta t}=8$ in
Fig.~\ref{fe2}.
For initial $e$ larger than those used in Figs.~\ref{fe0123} and
\ref{fe2} (e.g., $e = 0.6$), $a$ and $\dot{\psi}_{\s P}/n$ can drop
initially as $e$ declines rapidly if $A_{\Delta t} \sim 10$.

\begin{figure}
\epsscale{0.7}
\plotone{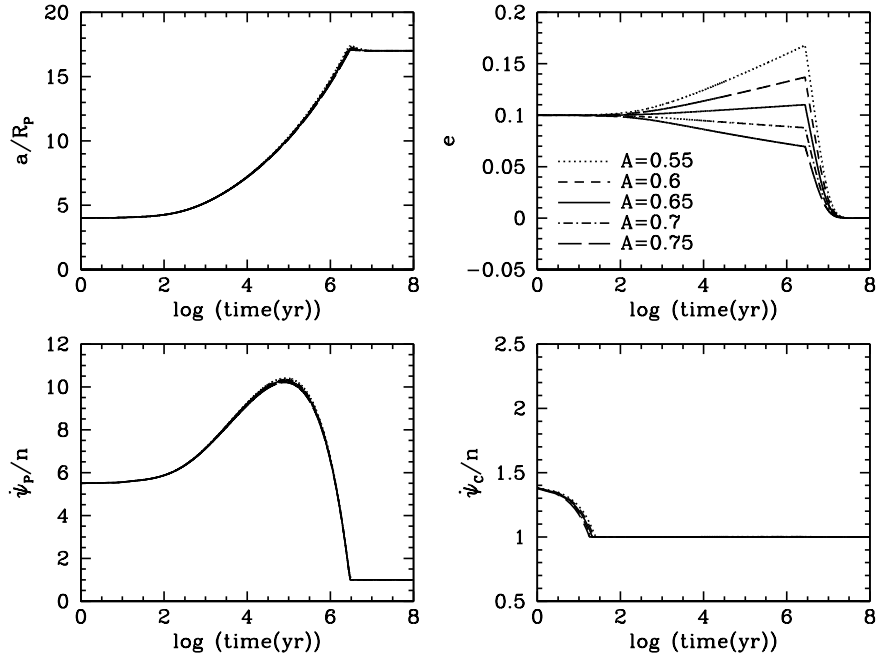}
\caption[Evolution in the constant $Q$ model with initial
$e=0.1$]{Evolution in the constant $Q$ model with initial $e=0.1$ and
$A_Q=0.55 - 0.75$.
The lines are almost identical for $a/R_{\s P}$, $\dot\psi_{\s P}/n$,
and $\dot\psi_{\s C}/n$.}
\label{qe1}
\end{figure}

Fig.~\ref{qe1} shows the evolution in the constant $Q$ model with
initial $e=0.1$, $\dot{\psi}_{\s C}=1.4n$ and a range of $A_Q$.
We choose this initial spin of Charon so that the $3n/2$ discontinuity
is avoided.
As long as no discontinuity is crossed, the coefficients in the tidal
evolution equations (Eqs.~[\ref{qdotpsi}]--[\ref{qedot}]) remain
constant and the eccentricity evolution is exponential.
Hence the lines in the $e$ versus $\log(t)$ graph look straight before
the spin of Pluto drops to below $3n/2$.
The spin of Charon stays slightly larger than synchronous, as
described by \citet{Greenberg84}, with the difference depending on
both $e$ and the smoothing range $\epsilon$ (note from
Eq.~[\ref{qdotpsi}] that, without smoothing,
$\langle{d \dot{\psi}_i/dt}\rangle >0$ for $\dot\psi_i/n = 1$ and
$< 0$ for $1 < \dot\psi_i/n < 3/2$ and $e < 0.243$).
Same as in the constant $\Delta t$ model, a suitable value of $A_Q$
can be chosen to keep $e$ nearly constant.

\begin{figure}[t]
\epsscale{0.7}
\plotone{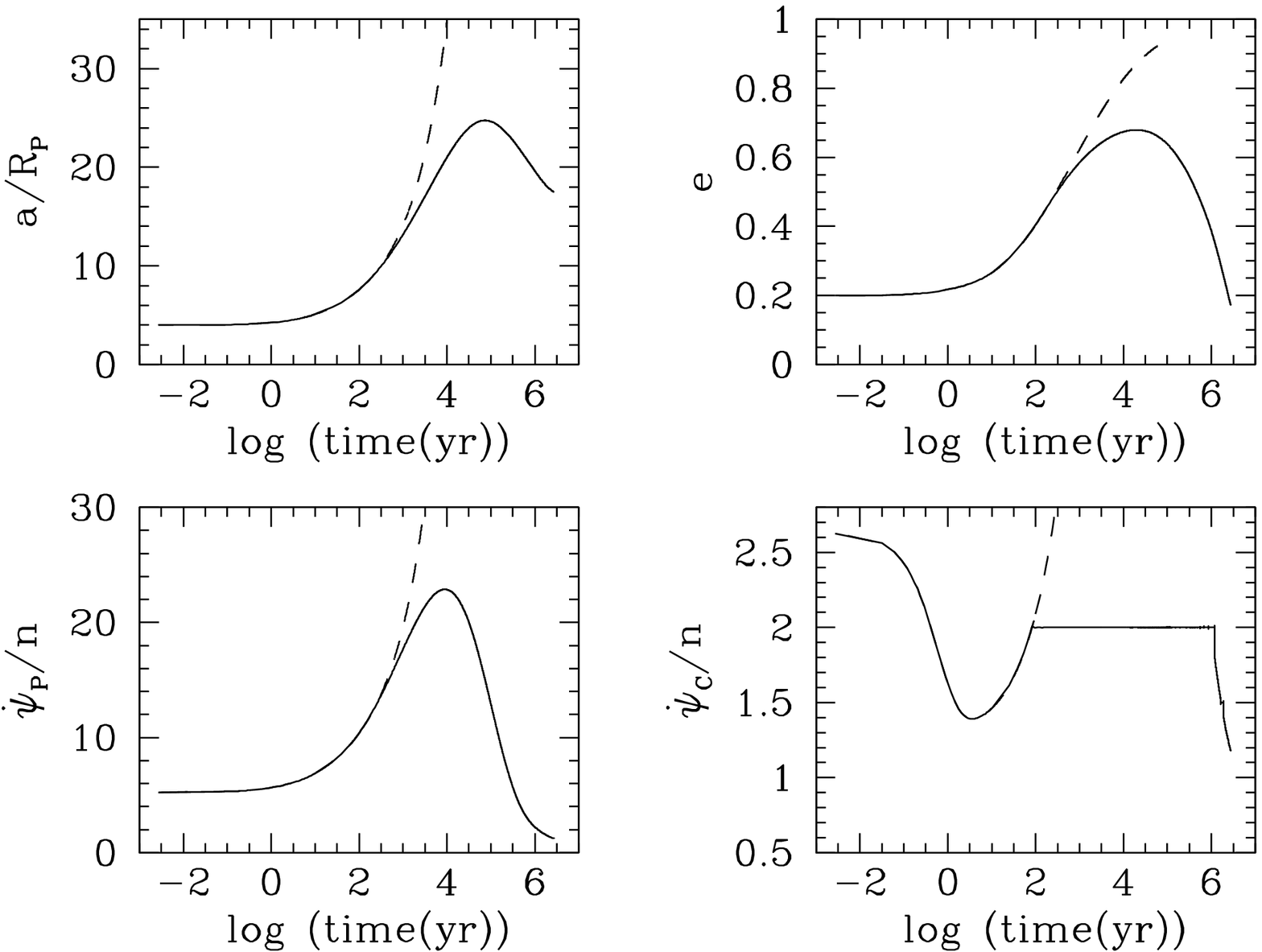}
\caption{Evolution in the constant $\Delta t$ model for
comparable tidal response and dissipation in both bodies
($A_{\Delta t} = 2.84$) and $C_{22i} = 0$ ({\it dashed lines}) or
$10^{-5}$ ({\it solid lines}).
Initial parameter values are $e = 0.2$, $\dot\psi_{\s P}/n = 5.28$,
and $\dot\psi_{\s C}/n = 2.63$.
\label{fig:A2.839}}
\end{figure}

In Section \ref{ICs} we show that $A_{\Delta t}$ and $A_Q \approx
0.5$--$3$ if Pluto and Charon have similar tidal response (in terms of
rigidities or Love numbers) and dissipation (in terms of $\Delta t$ or
$Q$).
For constant $Q$, this range includes the value ($A_Q \approx
0.65$--$1.15$) required to keep $e$ nearly constant.
For constant $\Delta t$, this range is significantly below the value
($A_{\Delta t} \approx 10$) required to keep $e$ nearly constant.
Fig.~\ref{fig:A2.839} shows the evolution for $A_{\Delta t} = 2.84$.
We include integrations for $C_{22{\s P}} = C_{22{\s C}} = 0$ and
$10^{-5}$ using the Bulirsch-Stoer code.
For the case of axial symmetry for both bodies ($C_{22{\s P}} =
C_{22{\s C}} = 0$), the dashed curves in Fig. \ref{fig:A2.839} show
that $e$ approaches 1, while the growth in $a$ well beyond the current
value drives $\dot\psi_{\s C}/n$ to values far above the 2:1
spin-orbit resonance and $\dot\psi_{\s P}/n$ to $> 30$.
The system can become unstable if the apoapse distance approaches
Pluto's Hill sphere radius.
The existence of the additional satellites Nix, Hydra, Keberos, and
Styx preclude even the large values of eccentricity on the way to
stable equilibrium in the case of axial asymmetry, if these satellites
were in orbit prior to the tidal expansion of Charon's orbit.
For this reason, we have considered larger values of $A_{\Delta t}$
that keep the value of $e$ at reasonably low values.
The case with permanent quadrupole moments ($C_{22{\s P}} =
C_{22{\s C}} = 10^{-5}$) in Fig.~\ref{fig:A2.839} is discussed in the
next subsection.

\subsection{Tidal Evolution with Non-zero $J_{2P}$ or $C_{22}$}
\label{results2}

In this subsection, we examine the effects of $J_{2P}$ and $C_{22i}$
on the tidal evolution of Pluto-Charon using results from $N$-body
integrations.

\begin{figure}[t]
\epsscale{0.7}
\plotone{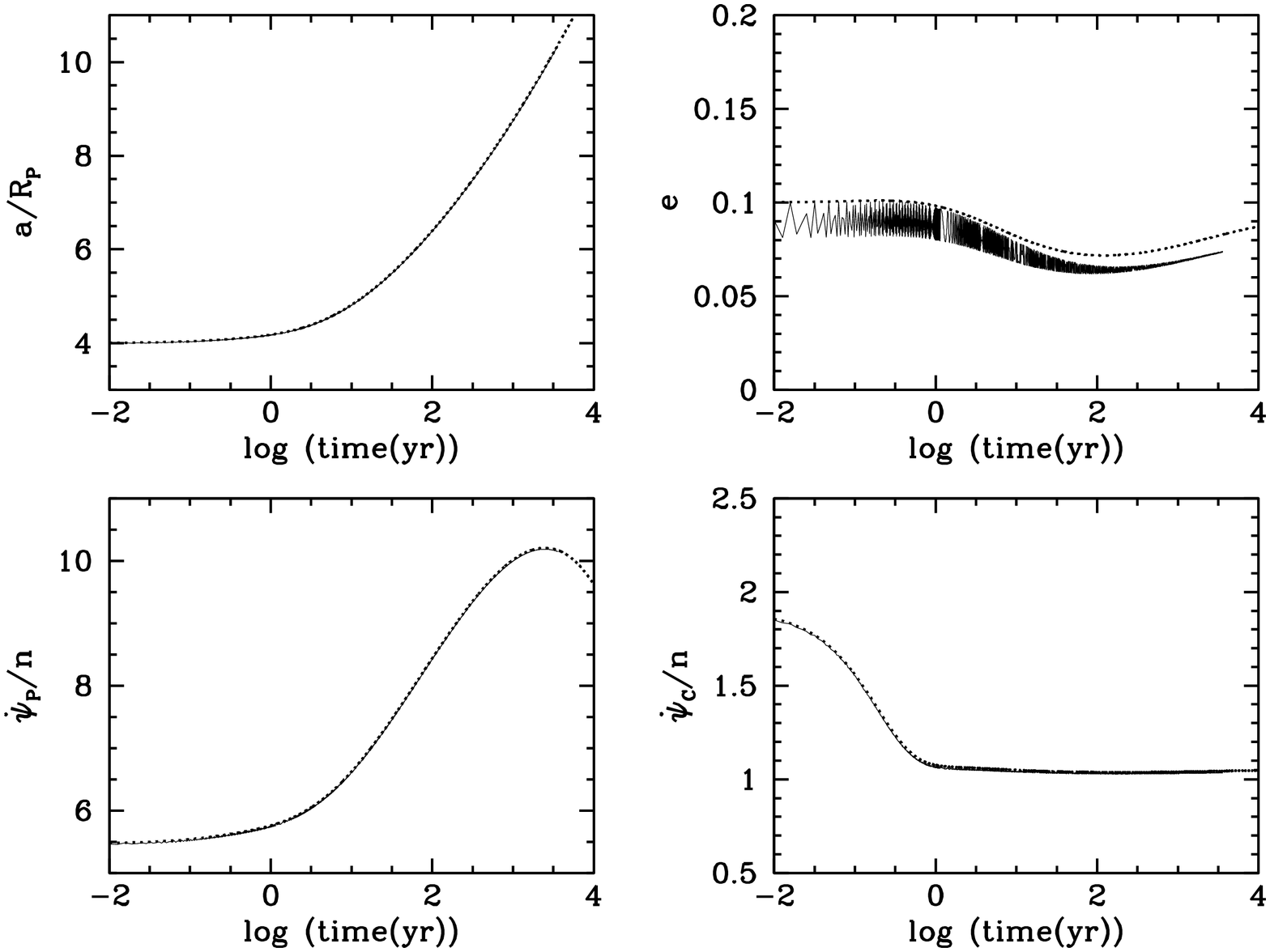}
\caption[Effects of $J_2$ on orbital evolution in the
constant $\Delta t$ model]{Effects of $J_2$ on orbital evolution in
the constant $\Delta t$ model. Dotted lines for $J_{2P}=0$ and solid
lines for $J_{2P}=0.1$ initially and decreasing with $\dot\psi_{\s P}^2$
(the lines are indistinguishable for $a/R_{\s P}$, $\dot\psi_{\s P}/n$,
and $\dot\psi_{\s C}/n$). Initial $e=0.1$ and $A_{\Delta t}=11$.}
\label{j2eff}
\end{figure}

Fig.~\ref{j2eff} shows the comparison between initial $J_{2P}=0$ and
$0.1$ evolution in the constant $\Delta t$ model,
with initial $e = 0.1$ and $A_{\Delta t} = 11$.
For the case with initial $J_{2P} = 0.1$, we assume that $J_{2P}$
decreases with $\dot\psi_{\s P}^2$ as in equation (\ref{eq:j2}).
Only the early stages of the evolution are affected by $J_{2{\s P}}$
when Charon is close, and we see in Fig. \ref{j2eff} that, with the
exception of the fluctuations in the osculating eccentricity, the
overall evolution is not that different from the case with
$J_{2{\s P}} = 0$.
In Section \ref{ICs} we estimate from Eq.~(\ref{j2meanmot}) that the
change of mean motion by $J_{2P}$ is $\sim 1\%$ initially.
Oscillations of $e$ in Fig.~\ref{j2eff} has initial amplitude $\approx
0.01$, which is about the order of the change of mean motion due to
$J_{2P}$.
Because the effect of $J_{2P}$ is relatively minor compared to that of
$C_{22i}$, we shall usually not include it in the evolutionary
calculations.

When $C_{22i}$ are non-zero, Charon can be captured into spin-orbit
resonances in both tidal models, which may result in very different
evolution when compared with the zero $C_{22i}$ cases.
The profound effect of non-zero $C_{22i}$  on the evolution is
demonstrated in Fig.~\ref{fig:A2.839}, where we compare integrations
with $C_{22{\s P}} = C_{22{\s C}} = 0$ and $10^{-5}$ for $A_{\Delta t}
= 2.84$.
For non-zero $C_{22i}$, Charon is captured into the 2:1 spin-orbit
resonance after having passed through this resonance once.
Charon gets a second passage through this commensurability when the
eccentricity grows sufficiently to raise the asymptotic or
pseudo-synchronous spin of Charon above that commensurability.
This capture suppresses the growth in eccentricity and allows the
evolution to proceed to the dual synchronous state, which does not
occur for $C_{22i} = 0$.
In this example, Charon's spin passes through the 3:2 spin-orbit
resonance three times without capture, except for a short time on its
last passage.
However, when we change the initial conditions by, e.g., changing the
initial directions that the long axes of Pluto and Charon are
pointing, we sometimes get long-term capture into the 3:2 resonance
instead.
This illustrates the probabilistic nature of such captures for large
eccentricity \citep{Goldreich66b}.

\begin{figure}
\epsscale{0.65}
\plotone{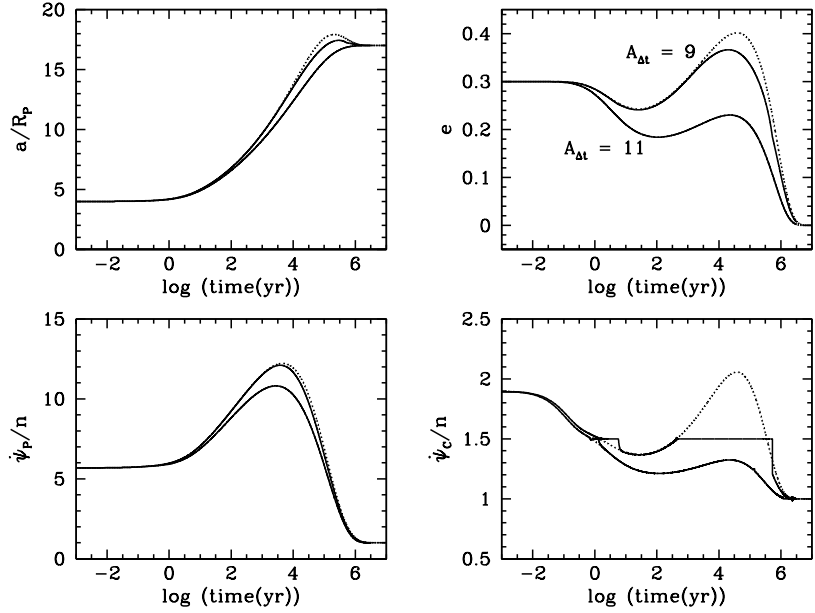}
\caption[Effects of $C_{22}$ on orbital evolution in the constant
$\Delta t$ model]{Effects of $C_{22}$ on orbital evolution in the
constant $\Delta t$ model. Dotted lines for $C_{22i} = 0$ and solid
lines for $C_{22i}=10^{-5}$. Initial $e=0.3$, and $A_{\Delta t} = 9$
(upper pair of lines in each panel) and 11 (lower pair of lines in
each panel, which are indistinguishable except for Charon's capture into
the 3:2 spin-orbit resonance when $e$ initially declines to
$0.285$).} \label{dte3c22}
\end{figure}

Fig.~\ref{dte3c22} shows the comparison between zero and non-zero
$C_{22i}$ evolution in the constant $\Delta t$ model for values of
$A_{\Delta t}$ that can keep $e$ roughly constant throughout most of
the evolution.
Initial $e=0.3$ and $A_{\Delta t} = 9$ (upper pair of lines in each
panels) and 11 (lower pair of lines in each panel).
For non-zero $C_{22i}$, Charon is caught in 3:2 spin-orbit resonance
when $e$ initially declines to $0.285$, where the asymptotic tidal
spin rate $\dot{\psi}_{\rm ps} = 1.5n$ (Eq.~[\ref{eq:psidotps}]),
but it escapes from the resonance when $e$ decreases to sufficiently
small values.
If $A_{\Delta t} = 11$, the relatively high dissipation in Charon
prevents the eccentricity from reaching $0.285$ again (and the
asymptotic value of $\dot\psi_{\s C}$ from reaching the 3:2 resonant
value), and the system proceeds normally to the dual synchronous
equilibrium state.
If $A_{\Delta t}=9$, $e$ rises above $0.285$ after the initial dip,
and the asymptotic tidal spin again goes to a value above the 3:2
spin-orbit resonance.
Capture into this resonance occurs at the second encounter where
Charon remains until the eccentricity again drops below stability and
Charon's spin goes directly to an asymptotic spin before capture into
the synchronous state.
Again the system proceeds to the dual synchronous equilibrium state.
The comparison with the $C_{22i} = 0$ calculations shows that the
capture into the 3:2 spin-orbit resonance happens to have only small
effects on the evolution of $a$, $e$, and $\dot\psi_{\s P}$ for the
examples in Fig.~\ref{dte3c22}.

\begin{figure}
\epsscale{0.9}
 \plottwo{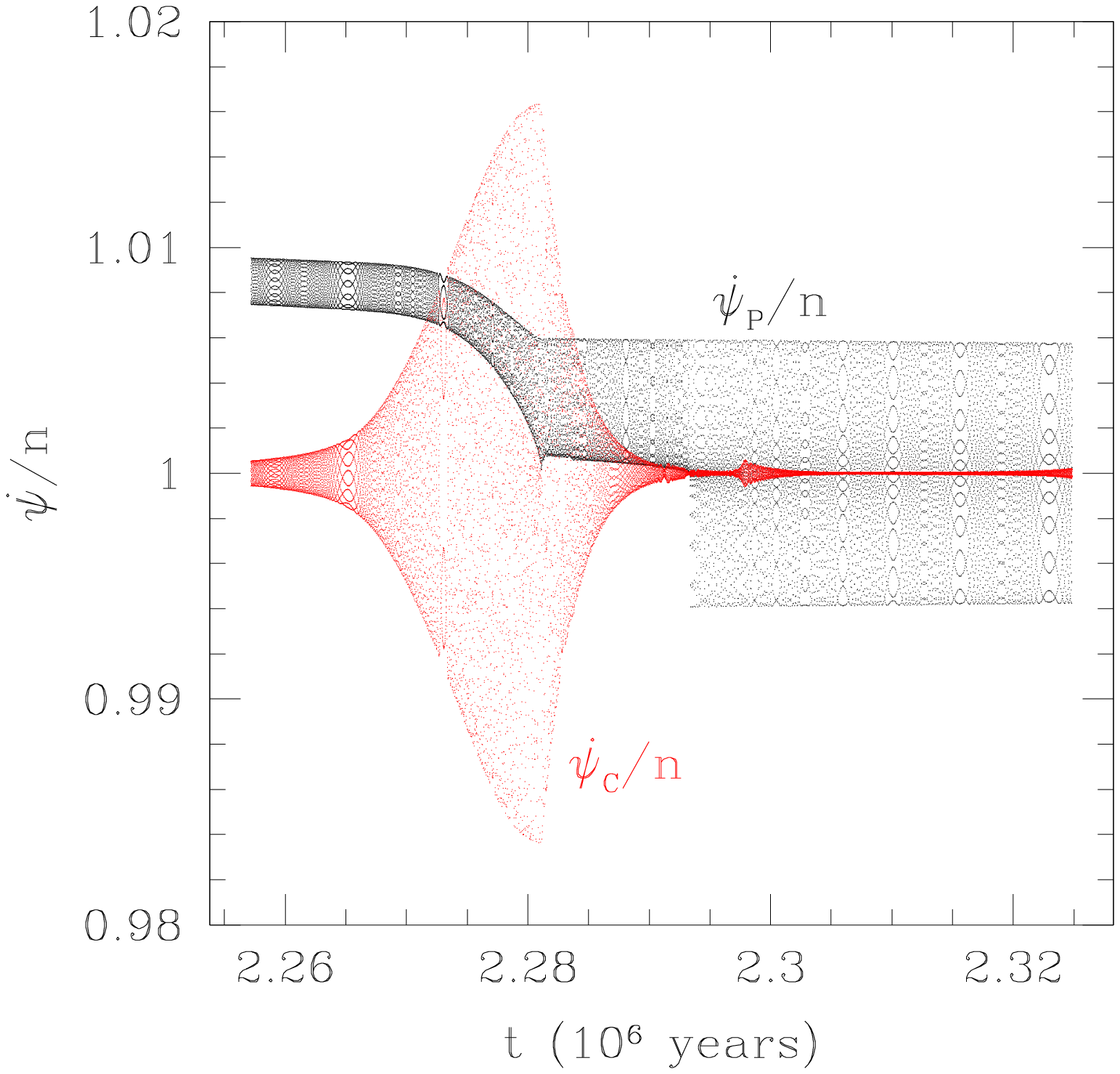}{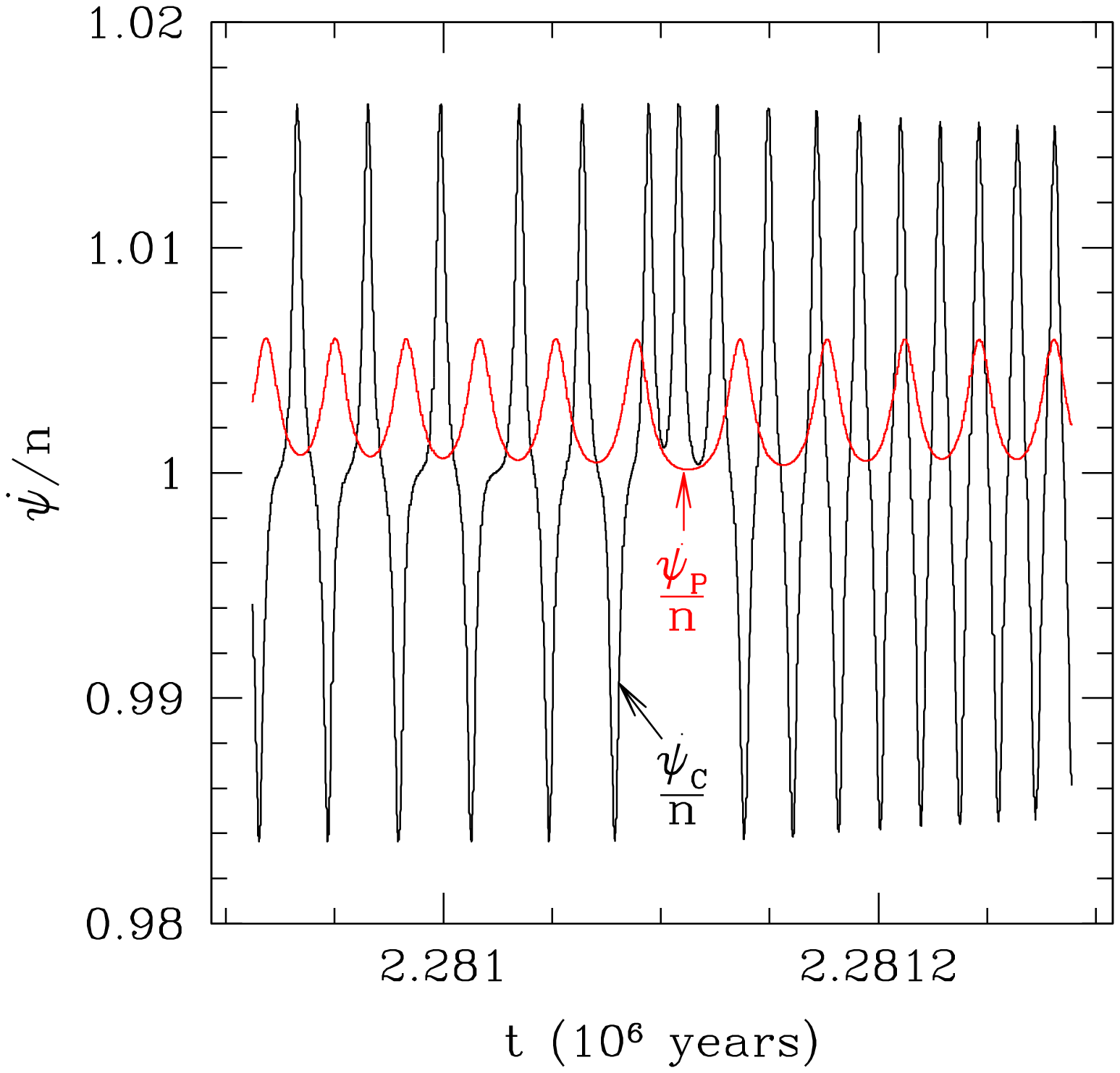}
\caption{Forced excitation of Charon's libration about synchronous
rotation from an interaction with Pluto's rotating figure as Pluto
approaches synchronous rotation.
Initial $e = 0.3$, $C_{22i} = 10^{-5}$, and $A_{\Delta t} = 9$.
The left panel shows the growth in amplitude of Charon's longitude
libration as Pluto approaches synchronous rotation, and the right
panel shows the transition of the variations in the spin rates as they
go out of resonance. \label{fig:resonance}} 
\end{figure}

For the cases with non-zero $C_{22i}$ shown in Fig.~\ref{dte3c22},
Charon reaches libration about synchronous rotation before Pluto, and
the librations rapidly damp to very small amplitude.
However, when Pluto approaches synchronous rotation, the
librations of Charon about the synchronous state are forced to
significant amplitude.
A blowup of these librations are shown in Fig.~\ref{fig:resonance} for
the case with $A_{\Delta t} = 9$, where the left panel shows
$\dot\psi_{\s C}/n$ and $\dot\psi_{\s P}/n$ as the latter transitions
into synchronous rotation.
The graphs are sparsely sampled to see both trends where
they are superposed.
The variations in the two spins are
anti-correlated during the rise in the amplitude of $\dot\psi_{\s C}$,
but that correlation is lost after passage over the peak as shown in
the right panel of Fig. \ref{fig:resonance}.
The periods of the
variation are no longer the same after the peak, and the tides rapidly
damp the amplitude of the libration of Charon.
It is not until the
amplitude is nearly zero that Pluto starts its libration about
synchronous rotation, which is damped to zero on a longer time
scale.
The period of free libration of Charon is approximately 345
days, and that of the variation in $\dot\psi_{\s C}$ is a little over
400 days.
We infer that it is the proximity of the forcing period from
the interaction with Pluto's axial asymmetry with the free period that
accounts for  the large growth in amplitude.
After the peak in the
amplitude, the period of Pluto's rotational variation changes and the
two variations are no longer near resonance. 

\begin{figure}[t]
\epsscale{0.7}
\plotone{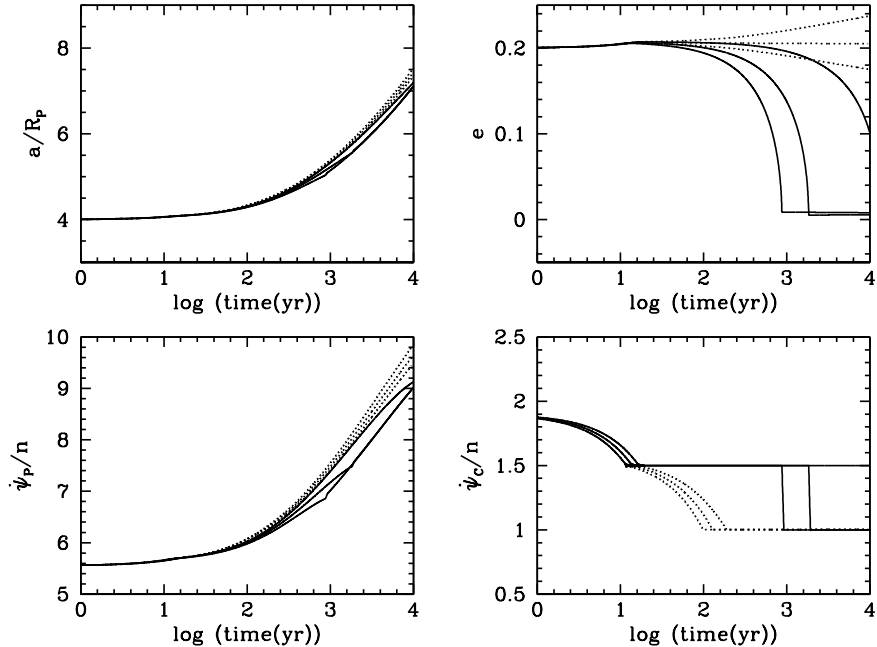}
\caption[Effects of $C_{22}$ on orbital evolution in the constant $Q$
model]{Effects of $C_{22}$ on orbital evolution in the constant $Q$
model. Dotted lines for $C_{22i} = 0$ and solid lines for $C_{22i} =
10^{-5}$. Initial $e=0.2$, $A_Q = 0.55$, 0.65, and 0.75 from top to
bottom in the eccentricity plot.} \label{qe2c22}
\end{figure}

Fig.~\ref{qe2c22} shows the comparison between zero and non-zero
$C_{22i}$ in the constant $Q$ model.
Initial $e=0.2$ and $A_Q=0.55$, $0.65$, and $0.75$ from top to bottom
in the eccentricity plot.
With non-zero $C_{22i}$, Charon is caught in the 3:2 spin-orbit
resonance in all three cases shown here.
The eccentricity drops quickly while Charon remains in the resonance,
which differs from the zero $C_{22i}$ cases where the eccentricity is
roughly constant up to $10^4\,$yr.
Charon's spin escapes from the resonance and reaches synchronous
rotation when $e$ becomes nearly zero.
The nearly zero $e$ also means that Pluto is not captured into any
spin-orbit resonance before synchronous rotation is reached (see
Section \ref{SOR}).

\begin{figure}[t]
\epsscale{0.7}
\plotone{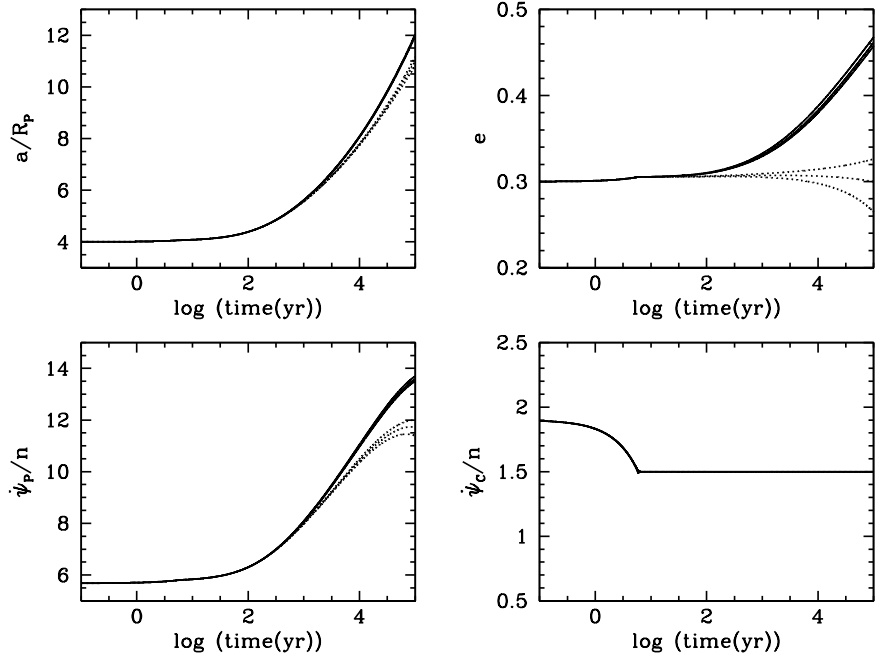}
\caption[Spin-orbit coupling in the constant $Q$ model with initial
$e=0.3$]{Effects of $C_{22}$ on orbital evolution in the constant $Q$
model with initial $e=0.3$. $C_{22i} = 0$ for dotted lines and
$10^{-5}$ for solid lines. $A_Q=1.13$, 1.14, and 1.15 (lines from top to
bottom). The spin of Charon is almost the same in all cases.}
\label{qe3c22}
\end{figure}

As the asymptotic spin rate in the constant $Q$ model is $3n/2$ for $e
> 0.235$, the spin of Charon can stay at $3n/2$ independent of the
value of $C_{22i}$ if the initial $e$ is larger.
However, there are differences in the evolution if $C_{22i}$ is
non-zero and Charon is actually in the spin-orbit resonance.
Fig.~\ref{qe3c22} compares evolution with zero and non-zero $C_{22i}$
in the constant $Q$ model for initial $e=0.3$ and $A_Q=1.13$, 1.14,
and 1.15 (lines from top to bottom).
The evolution of Charon's spin is almost the same in all cases, but
non-zero $C_{22i}$ causes $e$ to increase quickly here, compared to
the cases shown in Fig.~\ref{qe2c22}, where $e$ decreases.
(We stopped the calculations with non-zero $C_{22i}$ in
Fig.~\ref{qe3c22} at $t \approx 10^5\,$yr, because the evolution
equations for constant $Q$ are qualitatively inaccurate for $e \ga
0.36$.)
The value of $A_Q$ that keeps $e$ roughly constant would change when
$C_{22i}$ is non-zero.
We find that it is still possible to get a roughly constant $e$ by
using a smaller $A_Q$ for initial $e=0.2$, and larger $A_Q$ for
initial $e=0.3$.

\section{DISCUSSION}
\label{discussion}

\subsection{Tidal Evolution of Pluto-Charon}

We discuss in this subsection a number of issues concerning the tidal
evolution of Pluto-Charon, including the change of parameters, the use
of expanded equations for the constant $Q$ model, and the values of
$A$ in the two tidal models.

In Section \ref{results}, all results are generated using
$\bar{\mathcal{C}}_{\s P}=0.328$ and initial $a=4R_{\s P}$.
We have also performed calculations with other values of
$\bar{\mathcal{C}}_{\s P}$ and initial $a$, and find that they do not
affect our results qualitatively.
In particular, it is always possible to find $A$ that keeps $e$
roughly constant throughout most of the evolution.

\begin{figure}[t]
\epsscale{0.7}
\plotone{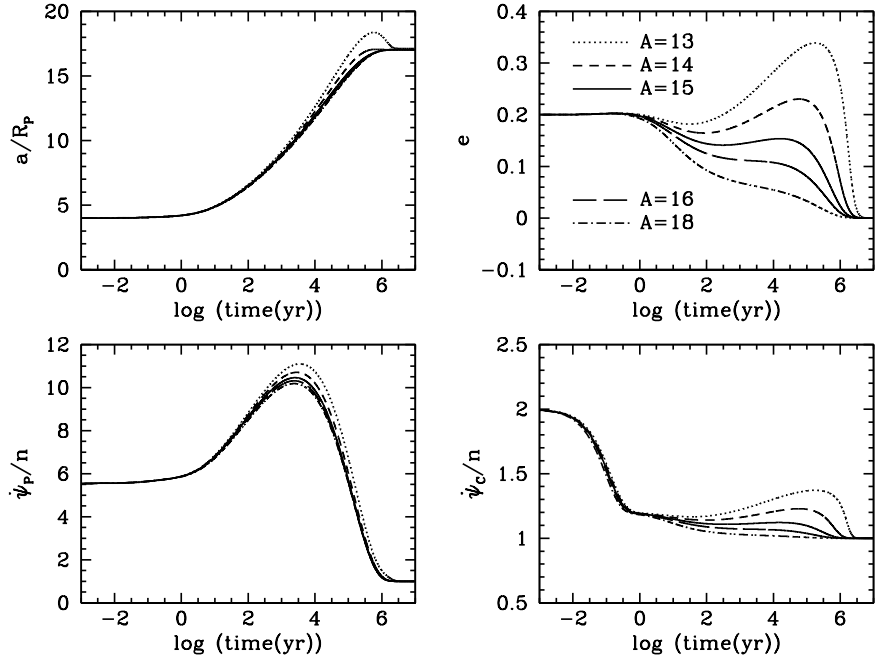}
\caption[Evolution in constant $\Delta t$ expanded equations with
initial $e=0.2$]{Evolution using constant $\Delta t$ expanded
equations with initial $e = 0.2$, $C_{22i} = 0$, and $A_{\Delta t} =
13$--$18$.}
\label{te2}
\end{figure}

Evolution equations are available in both closed form
(Eqs.~[\ref{fdotpsi}]--[\ref{fedot}]) and lowest order expansion in
$e$ (Eqs.~[\ref{tdotpsi}]--[\ref{tedot}]) in the constant $\Delta t$
model, while only expanded equations can be obtained for constant $Q$.
We compare the results from the two constant $\Delta t$ models to
give us an idea how good the results from the $O\left(e^2\right)$
equations of the constant $Q$ model are.
Fig.~\ref{te2} shows the evolution using the constant $\Delta t$
expanded equations, with the same initial conditions as in
Fig.~\ref{fe2}.
We find that similar eccentricity evolution can be recovered with the
expanded equations by increasing $A_{\Delta t}$.
For example, $e$ can be kept more or less constant with $A_{\Delta t}
= 14$ in Fig.~\ref{te2}, compared to $A_{\Delta t} = 10$ in
Fig.~\ref{fe2}.
Larger $A_Q$ is also required to keep $e$ more or less constant for
larger $e$ in the constant $Q$ model (see, e.g., Fig.~\ref{qe0123}),
which may be a result of truncating the higher order terms in $e$ in
the evolution equations.
Note, however, that the expanded equations for constant $Q$ are
qualitatively inaccurate for $e \ga 0.36$, because they do not give
the next discontinuous jump in the asymptotic spin from $3n/2$ to $2n$
(see Section \ref{Qmodel}).

The hypothesis that the small satellites, Nix, Hydra, Keberos, and
Styx, were brought to their current orbits by mean-motion resonances
with Charon is motivated by finding them currently near the 4:1, 6:1,
5:1, and 3:1 mean-motion commensurabilities with Charon, respectively.
As we show in Section \ref{results}, $a$ would overshoot the current
value if $e$ is large when $a$ reaches this value, and decays back to
the current value when $e$ decays.
The overshoot poses a problem for the resonant migration hypothesis,
as mean-motion resonances may not be sustained with decreasing $a$ of
Charon.

Our results show that both tidal models can keep the eccentricity of
Charon's orbit more or less constant during most of the evolution, but
that the values of $A$ needed differ by an order of magnitude: $A_Q
\approx 0.65$--$1.15$ and $A_{\Delta t} \approx 10$.
We also show in Fig.~\ref{fig:A2.839} that $A_{\Delta t} \approx 3$
would result in unacceptably large eccentricity and growth in $a$ well
beyond the current value (especially if $C_{22i} = 0$).
For both tidal models, we expect $A \approx 0.5$--$3$ if Pluto and
Charon have similar tidal response and dissipation.
While the values of $A_Q$ needed for keeping $e$ roughly constant are
reasonable, it is unclear that $A_{\Delta t} \approx 10$ can be
achieved with different assumptions about Pluto and Charon.
Generally we expect the dissipation in Pluto would be larger than that
in Charon (and hence smaller $A$), if Charon comes off more or less
intact after the impact.
On the other hand, if the differentiated Pluto has fluid Love number
$k_{f{\s P}}\approx 1$ by analogy with the Earth, while Charon is
homogeneous with $k_{f{\s C}} = 3/2$, the resulting $A$ would be
increased by the same factor.
Note that $\dot{\psi}_{\s P}/n$ is large compared with
$\dot{\psi}_{\s C}/n$ until close to the end of tidal evolution and
the tidal frequencies on Pluto and Charon are different.
Unless $A \approx 10$ is plausible, our results suggest that the
frequency dependence of the dissipation function $Q$ of real solid
materials may be closer to $Q =$ constant than $Q \propto 1/f$ (or
constant $\Delta t$).

\subsection{Spin-orbit Resonance of Charon}
\label{SOR}

\begin{figure}
\epsscale{0.69}
\plotone{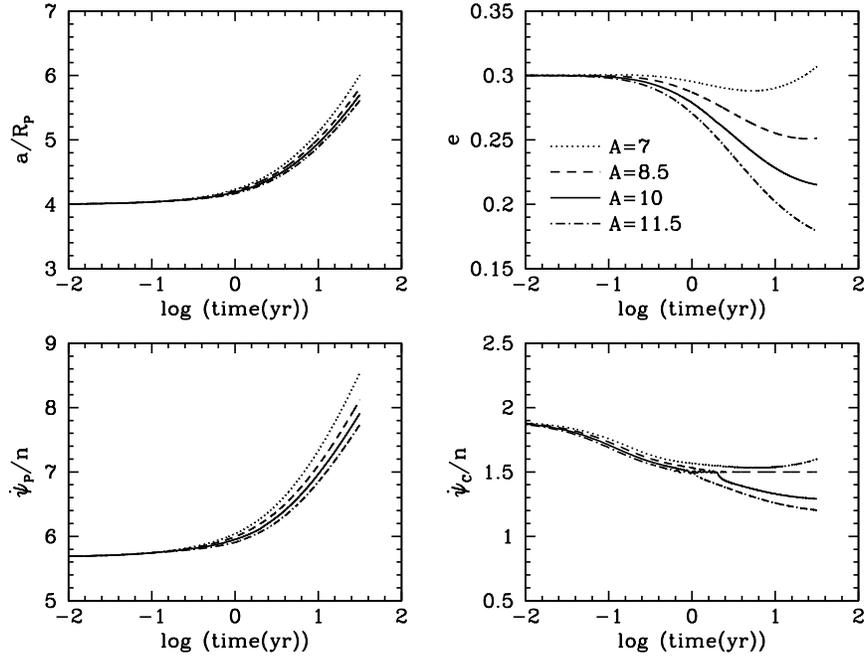}
\caption[Spin-orbit coupling in the constant $\Delta t$ model with a
range of $A_{\Delta t}$]{Spin-orbit coupling in the constant $\Delta
t$ model. Initial $\dot{\psi}_{\s C}=1.9n$ and $A_{\Delta t}=7$, 8.5, 10,
and 11.5 (lines from top to bottom).} \label{socccut}
\end{figure}

\begin{figure}
\epsscale{0.69}
\plotone{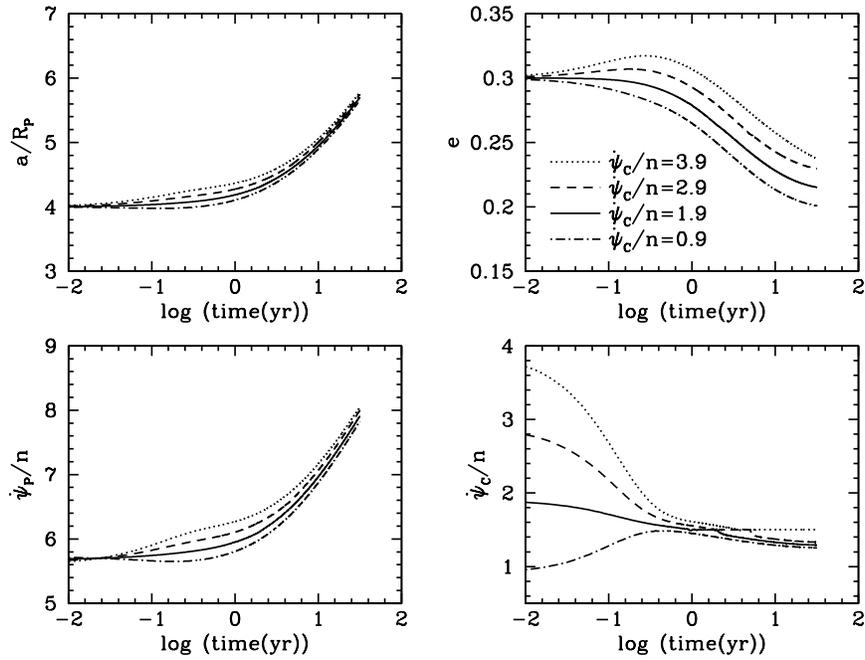}
\caption[Spin-orbit coupling in the constant $\Delta t$ model with a
range of initial $\dot{\psi}_{\s C}$]{Spin-orbit coupling in the constant
$\Delta t$ model. $A_{\Delta t}=10$ and initial $\dot{\psi}_{\s C}/n=3.9$,
2.9, 1.9, and 0.9 (lines from top to bottom).} \label{socAcut}
\end{figure}

\begin{figure}[htbp]
\epsscale{0.7}
\plotone{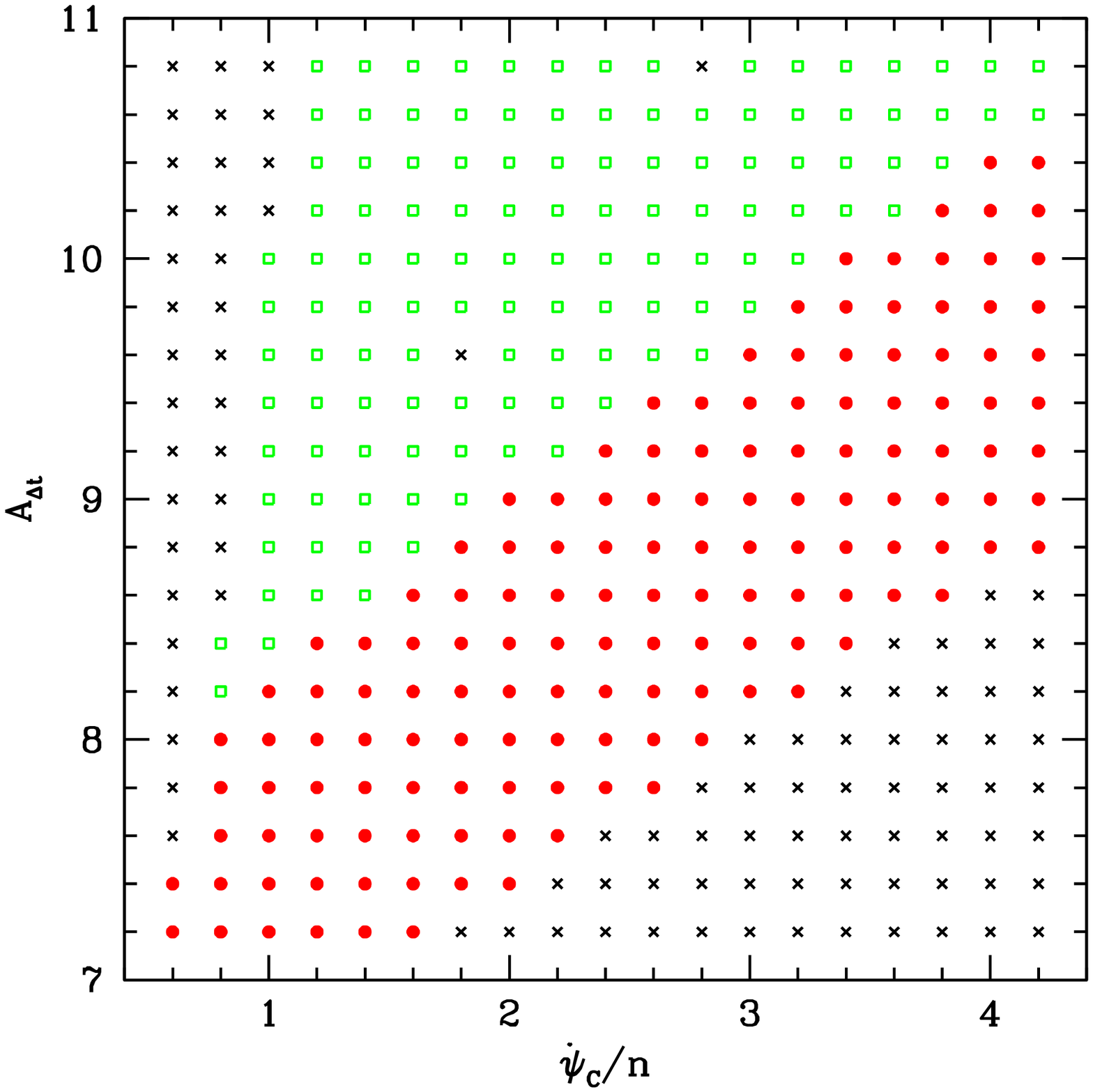}
\caption[Grid search of 3:2 spin-orbit resonance of Charon in the
constant $\Delta t$ model]{Grid search of 3:2 spin-orbit resonance of
Charon in the constant $\Delta t$ model. Initial $e=0.3$, and $A_{\Delta
t}$ and initial $\dot{\psi}_{\s C}/n$ are varied. Different symbols
stand for different results after $t\approx30$ years: The crosses
corresponds to conditions where Charon is not caught in the 3:2
resonance; the open squares where Charon is caught in the 3:2
resonance but escapes before $t\approx 30$ years; and the filled
circles where Charon remains in resonance up to $t\approx 30$ years.}
\label{socgrid}
\end{figure}

In this subsection we discuss in more detail the capture into and
escape from spin-orbit resonance for Charon when $C_{22i}$ are
non-zero.
We focus on the 3:2 spin-orbit resonance, because we are primarily
interested in evolution where $e$ does not become too large, and we
find only long-term capture into the 3:2 resonance if $e \la 0.36$.

We first consider the constant $\Delta t$ model, where
pseudo-synchronous spin of $3n/2$ corresponds to $e \approx 0.285$.
Since Charon's spin quickly reaches the pseudo-synchronous state, $e$
is usually close to the above value when it reaches $3n/2$.
Our $C_{22}=10^{-5}$ converts to $(\mathcal{B} - \mathcal{A})/\mathcal{C}
= 10^{-4}$ for homogeneous body.
According to Fig.~6 of \citet{Goldreich66b} for $(\mathcal{B} -
\mathcal{A})/\mathcal{C} = 10^{-4}$, the probability for Charon being
captured into the 3:2 resonance is close to unity at $e \approx
0.285$.
The range of $e$ for certain capture becomes narrower for smaller
$C_{22}$ (see their Fig.~7).

Fig.~\ref{socccut} shows the early evolution up to about $30$ years
in the constant $\Delta t$ model with initial $e=0.3$,
$\dot{\psi}_{\s C}=1.9n$, and a range of $A_{\Delta t}$.
We see that Charon does not reach the $3n/2$ spin if  $A_{\Delta t}$
is small ($\la 7$), and does not stay within the 3:2 resonance for
long if $e$ is damped for large $A_{\Delta t}$ ($\ga 10$).
Similarly, Fig.~\ref{socAcut} shows the early evolution with initial
$e = 0.3$, $A_{\Delta t} = 10$ and a range of initial
$\dot{\psi}_{\s C}$.
Charon does not reach the $3n/2$ spin if the initial
$\dot{\psi}_{\s C}$ is too small ($\la 1$), and the stability of the
resonance is maintained  if the initial $\dot{\psi}_{\s C}$ is large
($\ga 3.9$) and $e$ remains large.
The escape from 3:2 resonance at this stage does not preclude a
subsequent capture if $e$ rises above $0.285$ again (see, e.g.,
Fig.~\ref{dte3c22}).

Fig.~\ref{socgrid} shows a grid search for conditions under which
Charon stays within the 3:2 spin-orbit resonance in the constant
$\Delta t$ model.
In these runs, initial $e=0.3$ while $A_{\Delta t}$ and initial
$\dot{\psi}_{\s C}$ are varied.
The initial pointing directions of Pluto and Charon are randomly
chosen, and all runs end at $10^{9}$ seconds ($\approx 30$ years).
The crosses correspond to those conditions where Charon is not caught
in the 3:2 resonance.
Charon's spin never reaches $3n/2$ for the crosses in the upper left
and lower right corners, with initial $\dot{\psi}_{\s C}$ too low and
$e$ too large, respectively.
The open squares are those conditions where Charon is caught in the
3:2 resonance but escapes before $t\approx30$ years, while the filled
circles remain in resonance up to that time.
For the two crosses surrounded by open squares, it is possible to
get them caught in the 3:2 resonance by merely changing the initial
pointing directions of Pluto and Charon.

\begin{figure}
\epsscale{0.70}
\plotone{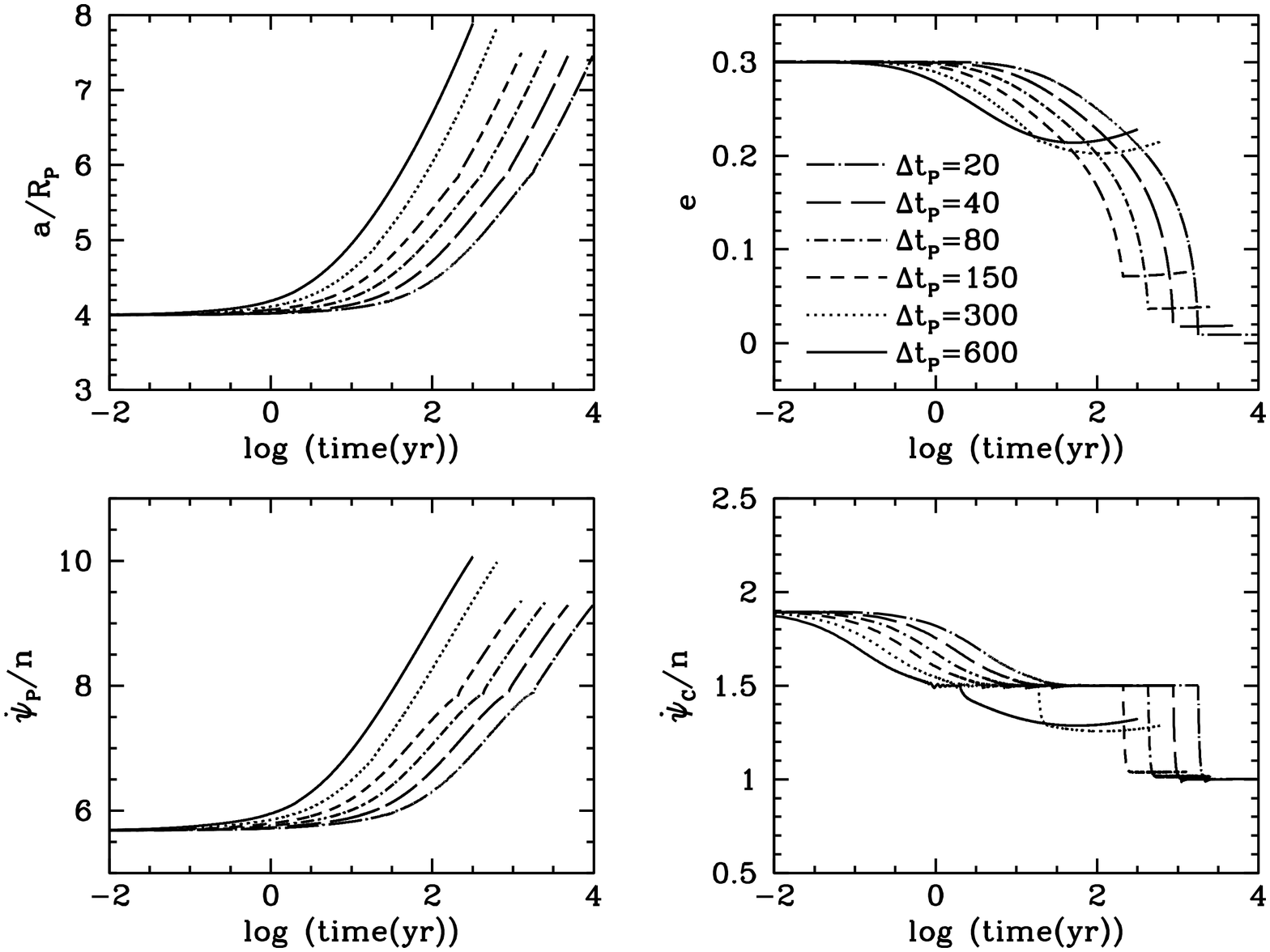}
\caption[Spin-orbit coupling in the constant $\Delta t$ model with a
range of $\Delta t_{\s P}$]{Spin-orbit coupling in the constant $\Delta t$
model. $A_{\Delta t}=10$ and $\Delta t_{\s P}=600$, 300, 150, 80, 40,
and 20 seconds (lines from top to bottom in the eccentricity plot and
from left to right in the other panels).}
\label{dt20-600}
\end{figure}

\begin{figure}
\epsscale{0.52}
\plotone{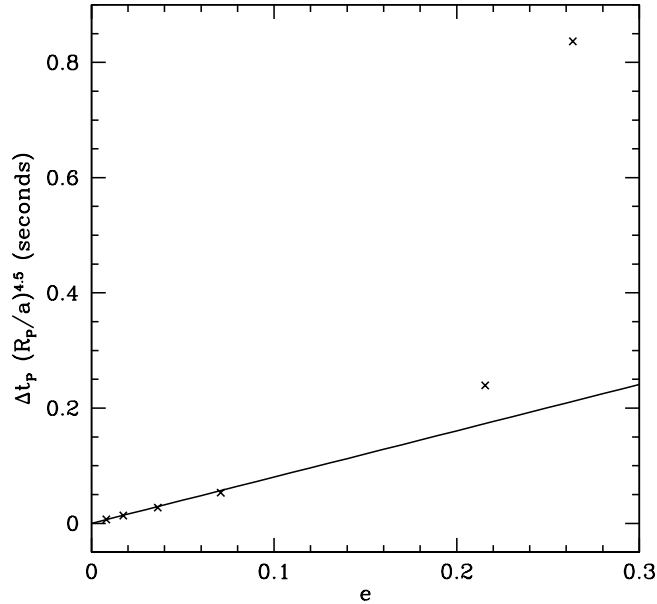}
\caption[Charon escapes from 3:2 spin-orbit resonance according to
stability limit]{$\Delta t_{\s P} (R_{\s P}/a)^{4.5}$ versus the
eccentricity at which Charon escapes from the 3:2 spin-orbit resonance
for the runs shown in Fig.~\ref{dt20-600}.
The straight line is the stability limit Eq.~(\ref{tsorcon}) at the
lowest order in $e$.}
\label{fc22condit}
\end{figure}

Charon escapes from the spin-orbit resonance if the tidal torque
exceeds the maximum possible restoring torque provided by $C_{22}$ on
the body.
By comparing the torques, the condition for the stability of the 3:2
spin-orbit resonance is (Eq.~[6] of \citealt{Goldreich66b}):
\begin{eqnarray}
\frac{k_{2C}}{C_{22C}} \frac{M_{\s P}^2}{M_{\s C} (M_{\s P}+M_{\s C})} \left(\frac{R_{\s C}}{a}
\right)^3 & < & \: \frac{14e}{n \Delta t_{\s C}} + O \left( e^3
\right), \label{tsorcon} \\ & \textrm{or} & \: 14e\ Q_{\s C}
+ O \left( e^3 \right) .
\end{eqnarray}
Eq.~(\ref{tsorcon}) can be rewritten in terms of $A_{\Delta t}$ and
Pluto's parameters.
Fig.~\ref{dt20-600} shows the evolution for a range of
$\Delta t_{\s P}$ and $A_{\Delta t}=10$.
Charon stays in the 3:2 resonance longer for smaller $\Delta t_{\s P}$.
To compare with the analytic stability condition, we plot
$\Delta t_{\s P} (R_{\s P}/a)^{4.5}$ against the eccentricity at which
Charon leaves the resonance in Fig.~\ref{fc22condit}, so that the $a$
dependence in Eq.~(\ref{tsorcon}) is removed.
The straight line shows Eq.~(\ref{tsorcon}) at the lowest order in
$e$.
The numerical results agree with the lowest-order analytic theory at
small $e$ but show departures at $e \ga 0.2$.

\begin{figure}[t]
\epsscale{0.7}
\plotone{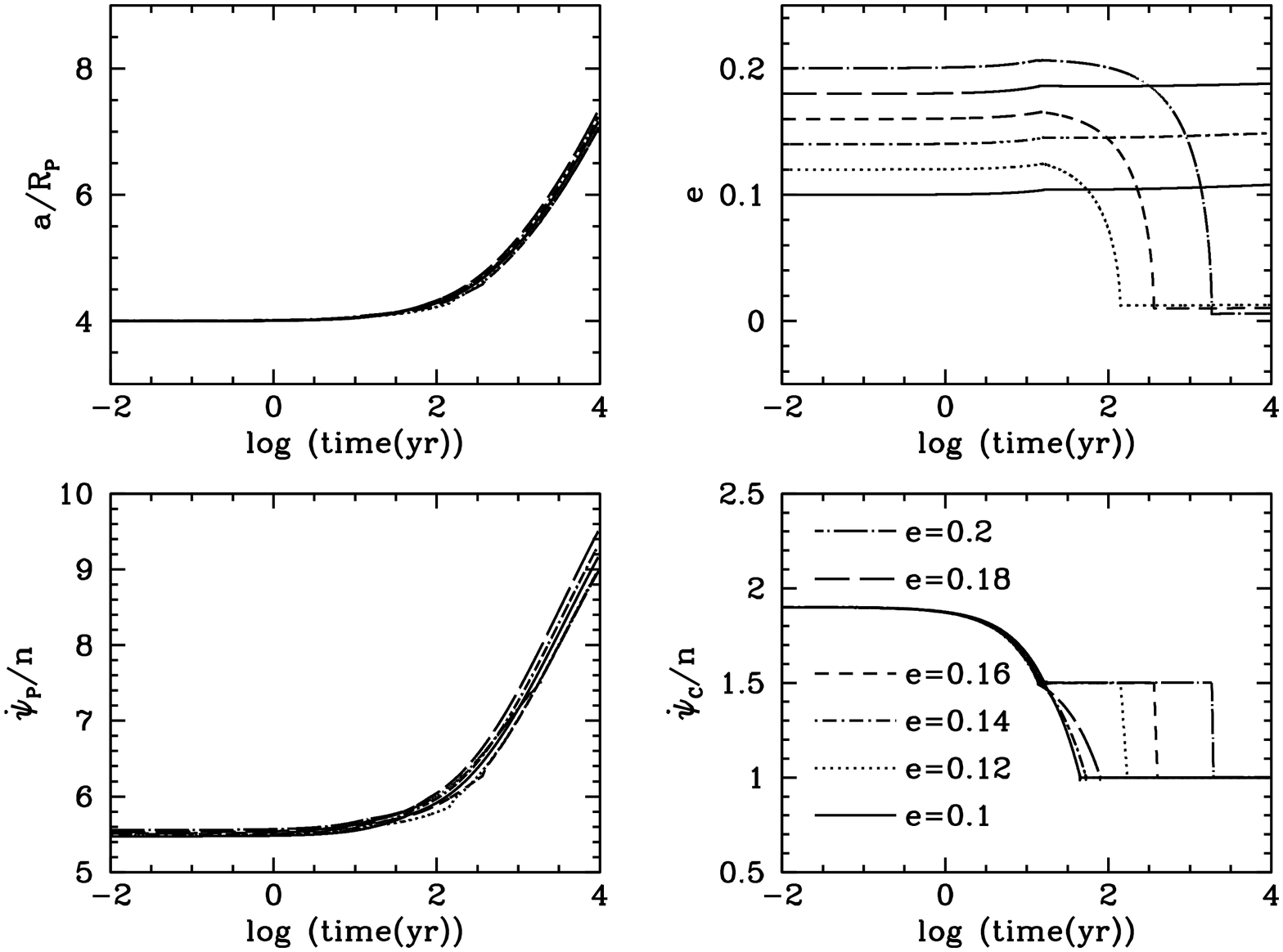}
\caption[Spin-orbit coupling in the constant $Q$ model with a range of
initial $e$]{Spin-orbit coupling in the constant $Q$ model. $A_Q=0.65$
and initial $e=0.2$, 0.18, 0.16, 0.14, 0.12, and 0.1 (lines from top
to bottom).} \label{qe1-2}
\end{figure}

In the constant $Q$ model, the capture into 3:2 spin-orbit resonance
from faster spin is certain if $e > 0.235$ and probabilistic if $e <
0.235$ (see Fig.~14 of \citealt{Goldreich66b}).
Fig.~\ref{qe1-2} illustrates the probabilistic capture using evolution
for a range of initial $e$ (0.1--0.2) and $A_Q=0.65$.
Although all three cases shown in Fig.\ref{qe2c22} with non-zero
$C_{22i}$, initial $e = 0.2$, and $A_Q = 0.55$--$0.75$ show capture
into the 3:2 resonance, we do find probabilistic capture when we try
other values of $A_Q$.

Capture of Charon into spin-orbit resonances other than 3:2 is
possible.
Fig.~\ref{fig:A2.839} shows an example of capture into 2:1 in the
constant $\Delta t$ model, but it requires $e$ to exceed $\approx
0.39$.
We have also seen {\it temporary} captures into 5:4 spin-orbit
resonance in more than one case in the constant $\Delta t$ model.
We note that \cite{Celletti07} have also seen the 5:4 spin-orbit
resonance, and \cite{Rodriguez12} have seen the 4:3 resonance.
The occurrence of these resonances is unexpected from first-order
perturbation theory which gives resonances only at spin rates that are
half-integer multiples of the mean motion (e.g.,
\citealt{Goldreich66b}).
The 5:4 is a second-order resonance that appears in second-order
perturbation theory \citep{Flynn05}.

Unlike Charon, Pluto is not captured into any spin-orbit resonance
before reaching synchronous rotation in nearly all of our calculations
with non-zero $C_{22i}$.
Since ${\dot\psi}_{P}/n$ is typically $\sim 5$--$6$ initially and
rises to $> 10$ before falling, by the time ${\dot\psi}_{P}/n$ reaches
values like $3/2$, the eccentricity is usually below the value where
the asymptotic spin rate is $3n/2$, and the probability of capturing
Pluto into spin-orbit resonance is small.

In our analysis, we have neglected several effects that could change
the probabiliy of capture into various spin-orbit resonances for both
Charon and Pluto.
These include an alternative tidal dissipation model that combines the
Andrade and Maxwell rheological models (e.g., \citealt{Makarov12}),
core-mantle interactions if the core is liquid (e.g.,
\citealt{Peale77,Correia09}), and collisions (e.g.,
\citealt{Correia12}).
However, even without these additional effects, we already observe the
occurrence of such captures for Charon and their effects on the
evolutionary track of Pluto-Charon.

\section{CONCLUSIONS}
\label{conclusions}

We have investigated the tidal evolution of Pluto-Charon
on an eccentric orbit under two different tidal models: constant
$\Delta t$ and constant $Q$.
Our calculations show the complete tidal evolution of a system of two
solid bodies of comparable size, where the spin angular momentum of
the two bodies is initially comparable to the orbital angular momentum.
The deviation from axial symmetry has been included in tidal
evolution, and the back reaction on the orbit must be accounted for to
conserve angular momentum.
Capture into spin-orbit resonances can profoundly affect the tidal
evolution of the system.

Motivated by binary asteroids (including those with comparable masses
in dual synchronous state, like (69230) Hermes and (90) Antiope),
\cite{Taylor10,Taylor11} have studied the tidal evolution of two
bodies.
They included higher order terms in the tidal potential, but
considered only circular orbit and zero $C_{22}$.
\cite{Rodriguez12} have studied the tidal evolution of super-Earths
close to their host stars.
They considered non-zero $C_{22}$ and eccentric orbit, and also found
that capture into spin-orbit resonances can significantly affect the
tidal evolution of eccentricity.
However, they included only the tides raised by the star on the
planet.
Both of these studies used the constant $\Delta t$ model only.

The equations used in our study are derived from a variety of existing
sources, and we provide a more comprehensive listing of the
coefficients in the evolution equations of the constant $Q$ model,
which depend discontinuously on the spin rate.
In both models, we find the value and range of relative rates of
tidal dissipation in Charon to that in Pluto that would result in
roughly constant eccentricity during most of the evolution.
In the constant $\Delta t$ model, the results are valid for arbitrary
eccentricity, which is not true for constant $Q$ (where the results
are qualitatively inaccurate for $e \gtrsim 0.36$ due to the necessary
truncations in the evolution equations).
However, the constant $Q$ model requires a more reasonable relative
rate of dissipation between Pluto and Charon ($A_Q \sim 1$).

It was assumed in previous studies (e.g., DPH97) that Charon would
achieve synchronous rotation
quickly after its formation. We show that this is not the case
for Charon on an eccentric orbit. The asymptotic spin depends on both
the eccentricity and the assumed tidal model. While the inferred large
oblateness of Pluto gives no significant change to the evolution, it
is found that the capture into spin-orbit resonance of Charon for
non-zero values of $C_{22}$ can change the relative dissipation rate
that keeps the eccentricity more or less constant during most of the
evolution.
In some cases (e.g., if $A_{\Delta t} \sim 1$), spin-orbit resonance
can allow smooth evolution to the final state of dual synchronous
rotation, whereas very large eccentricity and semimajor axis would
otherwise occur (which could lead to instability).
The conditions of capture into and escape from the 3:2 spin-orbit
resonance as a function of the orbital eccentricity agree with the
existing results in the literature.

\acknowledgments
The authors are grateful for the support of a Postgraduate Studentship
at the University of Hong Kong (WHC), Hong Kong RGC Grant HKU 7024/08P
(WHC and MHL), and the NASA Planetary Geology and Geophysics Program
under Grant NNX08AL76G (SJP).
We thank Robin Canup for useful discussions and the referees for their
helpful comments on the manuscript.

\bibliography{pluto}

\begin{thebibliography}{61}
\expandafter\ifx\csname natexlab\endcsname\relax\def\natexlab#1{#1}\fi
\expandafter\ifx\csname url\endcsname\relax
  \def\url#1{\texttt{#1}}\fi
\expandafter\ifx\csname urlprefix\endcsname\relax\def\urlprefix{URL }\fi
\providecommand{\eprint}[2][]{\url{#2}}
\providecommand{\bibinfo}[2]{#2}
\ifx\xfnm\relax \def\xfnm[#1]{\unskip,\space#1}\fi
%Type = Article
\bibitem[{{Alexander}(1973)}]{Alexander73}
\bibinfo{author}{{Alexander}, M.E.}, \bibinfo{year}{1973}.
\newblock \bibinfo{title}{{The weak friction approximation and tidal evolution
  in close binary systems}}.
\newblock \bibinfo{journal}{Ap\&SS} \bibinfo{volume}{23},
  \bibinfo{pages}{459--510}.
%Type = Inproceedings
\bibitem[{{Boss} and {Peale}(1986)}]{Boss86}
\bibinfo{author}{{Boss}, A.P.}, \bibinfo{author}{{Peale}, S.J.},
  \bibinfo{year}{1986}.
\newblock \bibinfo{title}{{Dynamical constraints on the origin of the Moon}},
  in: \bibinfo{editor}{{W.~K.~Hartmann, R.~J.~Phillips, \& G.~J.~Taylor}}
  (Ed.), \bibinfo{booktitle}{Origin of the Moon}, pp. \bibinfo{pages}{59--101}.
%Type = Article
\bibitem[{{Buie} et~al.(2006){Buie}, {Grundy}, {Young}, {Young} and
  {Stern}}]{Buie06}
\bibinfo{author}{{Buie}, M.W.}, \bibinfo{author}{{Grundy}, W.M.},
  \bibinfo{author}{{Young}, E.F.}, \bibinfo{author}{{Young}, L.A.},
  \bibinfo{author}{{Stern}, S.A.}, \bibinfo{year}{2006}.
\newblock \bibinfo{title}{{Orbits and Photometry of Pluto's Satellites: Charon,
  S/2005 P1, and S/2005 P2}}.
\newblock \bibinfo{journal}{Astron. J.} \bibinfo{volume}{132},
  \bibinfo{pages}{290--298}.
%Type = Article
\bibitem[{{Buie} et~al.(2010){Buie}, {Grundy}, {Young}, {Young} and
  {Stern}}]{Buie10}
\bibinfo{author}{{Buie}, M.W.}, \bibinfo{author}{{Grundy}, W.M.},
  \bibinfo{author}{{Young}, E.F.}, \bibinfo{author}{{Young}, L.A.},
  \bibinfo{author}{{Stern}, S.A.}, \bibinfo{year}{2010}.
\newblock \bibinfo{title}{{Pluto and Charon with the Hubble Space Telescope. I.
  Monitoring Global Change and Improved Surface Properties from Light Curves}}.
\newblock \bibinfo{journal}{Astron. J.} \bibinfo{volume}{139},
  \bibinfo{pages}{1117--1127}.
%Type = Article
\bibitem[{{Buie} et~al.(2012){Buie}, {Tholen} and {Grundy}}]{Buie12}
\bibinfo{author}{{Buie}, M.W.}, \bibinfo{author}{{Tholen}, D.J.},
  \bibinfo{author}{{Grundy}, W.M.}, \bibinfo{year}{2012}.
\newblock \bibinfo{title}{{The orbit of Charon is circular}}.
\newblock \bibinfo{journal}{Astron. J.} \bibinfo{volume}{144},
  \bibinfo{pages}{15}.
%Type = Article
\bibitem[{{Buie} et~al.(1997){Buie}, {Tholen} and {Wasserman}}]{Buie97}
\bibinfo{author}{{Buie}, M.W.}, \bibinfo{author}{{Tholen}, D.J.},
  \bibinfo{author}{{Wasserman}, L.H.}, \bibinfo{year}{1997}.
\newblock \bibinfo{title}{{Separate Lightcurves of Pluto and Charon}}.
\newblock \bibinfo{journal}{Icarus} \bibinfo{volume}{125},
  \bibinfo{pages}{233--244}.
%Type = Article
\bibitem[{{Cameron} and {Ward}(1976)}]{Cameron76}
\bibinfo{author}{{Cameron}, A.G.W.}, \bibinfo{author}{{Ward}, W.R.},
  \bibinfo{year}{1976}.
\newblock \bibinfo{title}{{The origin of the Moon}}.
\newblock \bibinfo{journal}{Lunar and Planetary Institute Science Conference
  Abstracts} \bibinfo{volume}{7}, \bibinfo{pages}{120}.
%Type = Article
\bibitem[{{Canup}(2004)}]{Canup04}
\bibinfo{author}{{Canup}, R.M.}, \bibinfo{year}{2004}.
\newblock \bibinfo{title}{{Dynamics of Lunar Formation}}.
\newblock \bibinfo{journal}{Ann. Rev. Astron. Astrophys.} \bibinfo{volume}{42},
  \bibinfo{pages}{441--475}.
%Type = Article
\bibitem[{{Canup}(2005)}]{Canup05}
\bibinfo{author}{{Canup}, R.M.}, \bibinfo{year}{2005}.
\newblock \bibinfo{title}{{A giant impact origin of Pluto-Charon}}.
\newblock \bibinfo{journal}{Science} \bibinfo{volume}{307},
  \bibinfo{pages}{546--550}.
%Type = Article
\bibitem[{{Castillo-Rogez} et~al.(2011){Castillo-Rogez}, {Efroimsky} and
  {Lainey}}]{Castillo11}
\bibinfo{author}{{Castillo-Rogez}, J.C.}, \bibinfo{author}{{Efroimsky}, M.},
  \bibinfo{author}{{Lainey}, V.}, \bibinfo{year}{2011}.
\newblock \bibinfo{title}{{The tidal history of Iapetus: Spin dynamics in the
  light of a refined dissipation model}}.
\newblock \bibinfo{journal}{J. Geophys. Res. (Planets)} \bibinfo{volume}{116},
  \bibinfo{pages}{9008}.
%Type = Article
\bibitem[{{Celletti} and {MacKay}(2007)}]{Celletti07}
\bibinfo{author}{{Celletti}, A.}, \bibinfo{author}{{MacKay}, R.},
  \bibinfo{year}{2007}.
\newblock \bibinfo{title}{{Regions of nonexistence of invariant tori for
  spin-orbit models}}.
\newblock \bibinfo{journal}{Chaos} \bibinfo{volume}{17},
  \bibinfo{pages}{043119}.
%Type = Article
\bibitem[{{Correia} and {Laskar}(2009)}]{Correia09}
\bibinfo{author}{{Correia}, A.C.M.}, \bibinfo{author}{{Laskar}, J.},
  \bibinfo{year}{2009}.
\newblock \bibinfo{title}{{Mercury's capture into the 3/2 spin-orbit resonance
  including the effect of core-mantle friction}}.
\newblock \bibinfo{journal}{Icarus} \bibinfo{volume}{201},
  \bibinfo{pages}{1--11}.
%Type = Article
\bibitem[{{Correia} and {Laskar}(2012)}]{Correia12}
\bibinfo{author}{{Correia}, A.C.M.}, \bibinfo{author}{{Laskar}, J.},
  \bibinfo{year}{2012}.
\newblock \bibinfo{title}{{Impact Cratering on Mercury: Consequences for the
  Spin Evolution}}.
\newblock \bibinfo{journal}{Astrophys. J.} \bibinfo{volume}{751},
  \bibinfo{pages}{L43}.
%Type = Incollection
\bibitem[{{Dobrovolskis} et~al.(1997){Dobrovolskis}, {Peale} and
  {Harris}}]{Dobrovolskis97}
\bibinfo{author}{{Dobrovolskis}, A.R.}, \bibinfo{author}{{Peale}, S.J.},
  \bibinfo{author}{{Harris}, A.W.}, \bibinfo{year}{1997}.
\newblock \bibinfo{title}{{Dynamics of the Pluto-Charon binary}}, in:
  \bibinfo{editor}{{S.~A.~Stern, \& D.~J. Tholen}} (Ed.),
  \bibinfo{booktitle}{Pluto and Charon}. \bibinfo{publisher}{Univ. Arizona
  Press}, \bibinfo{address}{Tucson, AZ}, p. \bibinfo{pages}{159}.
%Type = Article
\bibitem[{{Efroimsky} and {Williams}(2009)}]{Efroimsky09}
\bibinfo{author}{{Efroimsky}, M.}, \bibinfo{author}{{Williams}, J.G.},
  \bibinfo{year}{2009}.
\newblock \bibinfo{title}{{Tidal torques: a critical review of some
  techniques}}.
\newblock \bibinfo{journal}{Celest. Mech. Dyn. Astron.} \bibinfo{volume}{104},
  \bibinfo{pages}{257--289}.
%Type = Article
\bibitem[{{Farinella} et~al.(1979){Farinella}, {Milani}, {Nobili} and
  {Valsecchi}}]{Farinella79}
\bibinfo{author}{{Farinella}, P.}, \bibinfo{author}{{Milani}, A.},
  \bibinfo{author}{{Nobili}, A.M.}, \bibinfo{author}{{Valsecchi}, G.B.},
  \bibinfo{year}{1979}.
\newblock \bibinfo{title}{{Tidal evolution and the Pluto-Charon system}}.
\newblock \bibinfo{journal}{Moon Planets} \bibinfo{volume}{20},
  \bibinfo{pages}{415--421}.
%Type = Article
\bibitem[{{Ferraz-Mello} et~al.(2008){Ferraz-Mello}, {Rodr{\'{\i}}guez} and
  {Hussmann}}]{Ferraz-Mello08}
\bibinfo{author}{{Ferraz-Mello}, S.}, \bibinfo{author}{{Rodr{\'{\i}}guez}, A.},
  \bibinfo{author}{{Hussmann}, H.}, \bibinfo{year}{2008}.
\newblock \bibinfo{title}{{Tidal friction in close-in satellites and
  exoplanets: The Darwin theory re-visited}}.
\newblock \bibinfo{journal}{Celest. Mech. Dyn. Astron.} \bibinfo{volume}{101},
  \bibinfo{pages}{171--201}.
%Type = Article
\bibitem[{{Flynn} and {Saha}(2005)}]{Flynn05}
\bibinfo{author}{{Flynn}, A.E.}, \bibinfo{author}{{Saha}, P.},
  \bibinfo{year}{2005}.
\newblock \bibinfo{title}{{Second-Order Perturbation Theory for Spin-Orbit
  Resonances}}.
\newblock \bibinfo{journal}{Astron. J.} \bibinfo{volume}{130},
  \bibinfo{pages}{295--307}.
%Type = Article
\bibitem[{{Gerstenkorn}(1955)}]{Gerstenkorn55}
\bibinfo{author}{{Gerstenkorn}, H.}, \bibinfo{year}{1955}.
\newblock \bibinfo{title}{{{\"U}ber Gezeitenreibung beim Zweik{\"o}rperproblem.
  Mit 4 Textabbildungen}}.
\newblock \bibinfo{journal}{Zeitschrift f{\"u}r Astrophysik}
  \bibinfo{volume}{36}, \bibinfo{pages}{245}.
%Type = Article
\bibitem[{{Goldreich}(1966)}]{Goldreich66a}
\bibinfo{author}{{Goldreich}, P.}, \bibinfo{year}{1966}.
\newblock \bibinfo{title}{{Final spin states of planets and satellites}}.
\newblock \bibinfo{journal}{Astron. J.} \bibinfo{volume}{71},
  \bibinfo{pages}{1--7}.
%Type = Article
\bibitem[{{Goldreich} and {Peale}(1966)}]{Goldreich66b}
\bibinfo{author}{{Goldreich}, P.}, \bibinfo{author}{{Peale}, S.J.},
  \bibinfo{year}{1966}.
\newblock \bibinfo{title}{{Spin-orbit coupling in the solar system}}.
\newblock \bibinfo{journal}{Astron. J.} \bibinfo{volume}{71},
  \bibinfo{pages}{425--438}.
%Type = Article
\bibitem[{{Greenberg} and {Weidenschilling}(1984)}]{Greenberg84}
\bibinfo{author}{{Greenberg}, R.}, \bibinfo{author}{{Weidenschilling}, S.J.},
  \bibinfo{year}{1984}.
\newblock \bibinfo{title}{{How fast do Galilean satellites spin?}}
\newblock \bibinfo{journal}{Icarus} \bibinfo{volume}{58},
  \bibinfo{pages}{186--196}.
%Type = Article
\bibitem[{{Hut}(1981)}]{Hut81}
\bibinfo{author}{{Hut}, P.}, \bibinfo{year}{1981}.
\newblock \bibinfo{title}{{Tidal evolution in close binary systems}}.
\newblock \bibinfo{journal}{Astron. Astrophys.} \bibinfo{volume}{99},
  \bibinfo{pages}{126--140}.
%Type = Article
\bibitem[{{Kaula}(1964)}]{Kaula64}
\bibinfo{author}{{Kaula}, W.M.}, \bibinfo{year}{1964}.
\newblock \bibinfo{title}{{Tidal dissipation by solid friction and the
  resulting orbital evolution}}.
\newblock \bibinfo{journal}{Rev. Geophys. Space Phys.} \bibinfo{volume}{2},
  \bibinfo{pages}{661--685}.
%Type = Article
\bibitem[{{Lee} and {Peale}(2002)}]{Lee02}
\bibinfo{author}{{Lee}, M.H.}, \bibinfo{author}{{Peale}, S.J.},
  \bibinfo{year}{2002}.
\newblock \bibinfo{title}{{Dynamics and origin of the 2:1 orbital resonances of
  the GJ 876 planets}}.
\newblock \bibinfo{journal}{Astrophys. J.} \bibinfo{volume}{567},
  \bibinfo{pages}{596--609}.
%Type = Article
\bibitem[{{Levison} and {Duncan}(1994)}]{Levison94}
\bibinfo{author}{{Levison}, H.F.}, \bibinfo{author}{{Duncan}, M.J.},
  \bibinfo{year}{1994}.
\newblock \bibinfo{title}{{The long-term dynamical behavior of short-period
  comets}}.
\newblock \bibinfo{journal}{Icarus} \bibinfo{volume}{108},
  \bibinfo{pages}{18--36}.
%Type = Article
\bibitem[{{Lin}(1981)}]{Lin81}
\bibinfo{author}{{Lin}, D.N.C.}, \bibinfo{year}{1981}.
\newblock \bibinfo{title}{{On the origin of the Pluto-Charon system}}.
\newblock \bibinfo{journal}{Mon. Not. R. Astron. Soc.} \bibinfo{volume}{197},
  \bibinfo{pages}{1081--1085}.
%Type = Article
\bibitem[{{Lithwick} and {Wu}(2008)}]{Lithwick08a}
\bibinfo{author}{{Lithwick}, Y.}, \bibinfo{author}{{Wu}, Y.},
  \bibinfo{year}{2008}.
\newblock \bibinfo{title}{{On the Origin of Pluto's Minor Moons, Nix and
  Hydra}}.
\newblock \bibinfo{journal}{preprint (arXiv:0802.2951)} .
%Type = Article
\bibitem[{{Makarov} et~al.(2012){Makarov}, {Berghea} and
  {Efroimsky}}]{Makarov12}
\bibinfo{author}{{Makarov}, V.V.}, \bibinfo{author}{{Berghea}, C.},
  \bibinfo{author}{{Efroimsky}, M.}, \bibinfo{year}{2012}.
\newblock \bibinfo{title}{{Dynamical Evolution and Spin-Orbit Resonances of
  Potentially Habitable Exoplanets: The Case of GJ 581d}}.
\newblock \bibinfo{journal}{Astrophys. J.} \bibinfo{volume}{761},
  \bibinfo{pages}{83}.
%Type = Article
\bibitem[{{McKinnon}(1984)}]{McKinnon84}
\bibinfo{author}{{McKinnon}, W.B.}, \bibinfo{year}{1984}.
\newblock \bibinfo{title}{{On the origin of Triton and Pluto}}.
\newblock \bibinfo{journal}{Nature} \bibinfo{volume}{311},
  \bibinfo{pages}{355--358}.
%Type = Article
\bibitem[{{McKinnon}(1989)}]{McKinnon89}
\bibinfo{author}{{McKinnon}, W.B.}, \bibinfo{year}{1989}.
\newblock \bibinfo{title}{{On the origin of the Pluto-Charon binary}}.
\newblock \bibinfo{journal}{Astrophys. J.} \bibinfo{volume}{344},
  \bibinfo{pages}{L41--L44}.
%Type = Incollection
\bibitem[{{McKinnon} et~al.(2008){McKinnon}, {Prialnik}, {Stern} and
  {Coradini}}]{McKinnon08}
\bibinfo{author}{{McKinnon}, W.B.}, \bibinfo{author}{{Prialnik}, D.},
  \bibinfo{author}{{Stern}, S.A.}, \bibinfo{author}{{Coradini}, A.},
  \bibinfo{year}{2008}.
\newblock \bibinfo{title}{{Structure and evolution of Kuiper belt objects and
  dwarf planets}}, in: \bibinfo{editor}{{Barucci, M.~A., Boehnhardt, H.,
  Cruikshank, D.~P., \& Morbidelli, A. }} (Ed.), \bibinfo{booktitle}{The Solar
  System Beyond Neptune}. \bibinfo{publisher}{Univ. Arizona Press},
  \bibinfo{address}{Tucson, AZ}, p. \bibinfo{pages}{213}.
%Type = Article
\bibitem[{{Mignard}(1979)}]{Mignard79}
\bibinfo{author}{{Mignard}, F.}, \bibinfo{year}{1979}.
\newblock \bibinfo{title}{{The evolution of the lunar orbit revisited. I}}.
\newblock \bibinfo{journal}{Moon Planets} \bibinfo{volume}{20},
  \bibinfo{pages}{301--315}.
%Type = Article
\bibitem[{{Mignard}(1980)}]{Mignard80}
\bibinfo{author}{{Mignard}, F.}, \bibinfo{year}{1980}.
\newblock \bibinfo{title}{{The evolution of the lunar orbit revisited. II}}.
\newblock \bibinfo{journal}{Moon Planets} \bibinfo{volume}{23},
  \bibinfo{pages}{185--201}.
%Type = Article
\bibitem[{{Mignard}(1981a)}]{Mignard81a}
\bibinfo{author}{{Mignard}, F.}, \bibinfo{year}{1981}a.
\newblock \bibinfo{title}{{On a possible origin of Charon}}.
\newblock \bibinfo{journal}{Astron. Astrophys.} \bibinfo{volume}{96},
  \bibinfo{pages}{L1}.
%Type = Article
\bibitem[{{Mignard}(1981b)}]{Mignard81b}
\bibinfo{author}{{Mignard}, F.}, \bibinfo{year}{1981}b.
\newblock \bibinfo{title}{{The lunar orbit revisited. III}}.
\newblock \bibinfo{journal}{Moon Planets} \bibinfo{volume}{24},
  \bibinfo{pages}{189--207}.
%Type = Book
\bibitem[{{Munk} and {MacDonald}(1960)}]{Munk60}
\bibinfo{author}{{Munk}, W.H.}, \bibinfo{author}{{MacDonald}, G.J.F.},
  \bibinfo{year}{1960}.
\newblock \bibinfo{title}{{The Rotation of the Earth: A Geophysical
  Discussion}}.
\newblock \bibinfo{publisher}{Cambridge Univ. Press},
  \bibinfo{address}{London}.
%Type = Book
\bibitem[{{Murray} and {Dermott}(1999)}]{Murray99}
\bibinfo{author}{{Murray}, C.D.}, \bibinfo{author}{{Dermott}, S.F.},
  \bibinfo{year}{1999}.
\newblock \bibinfo{title}{{Solar System Dynamics}}.
\newblock \bibinfo{publisher}{Cambridge Univ. Press},
  \bibinfo{address}{Cambridge}.
%Type = Article
\bibitem[{{Peale}(1973)}]{Peale73}
\bibinfo{author}{{Peale}, S.J.}, \bibinfo{year}{1973}.
\newblock \bibinfo{title}{{Rotation of solid bodies in the solar system.}}
\newblock \bibinfo{journal}{Rev. Geophys. Space Phys.} \bibinfo{volume}{11},
  \bibinfo{pages}{767--793}.
%Type = Article
\bibitem[{{Peale}(1999)}]{Peale99}
\bibinfo{author}{{Peale}, S.J.}, \bibinfo{year}{1999}.
\newblock \bibinfo{title}{{Origin and evolution of the natural satellites}}.
\newblock \bibinfo{journal}{Ann. Rev. Astron. Astrophys.} \bibinfo{volume}{37},
  \bibinfo{pages}{533--602}.
%Type = Article
\bibitem[{{Peale}(2005)}]{Peale05}
\bibinfo{author}{{Peale}, S.J.}, \bibinfo{year}{2005}.
\newblock \bibinfo{title}{{The free precession and libration of Mercury}}.
\newblock \bibinfo{journal}{Icarus} \bibinfo{volume}{178},
  \bibinfo{pages}{4--18}.
%Type = Incollection
\bibitem[{{Peale}(2007)}]{Peale07}
\bibinfo{author}{{Peale}, S.J.}, \bibinfo{year}{2007}.
\newblock \bibinfo{title}{The origin of the natural satellites}, in:
  \bibinfo{editor}{{Spohn}, T.}, \bibinfo{editor}{{Schubert}, G.} (Eds.),
  \bibinfo{booktitle}{Treatise on Geophysics, Vol. 10, Planets and Moons}.
  \bibinfo{publisher}{Elsevier B.V.}, \bibinfo{address}{Amsterdam}, p.
  \bibinfo{pages}{465}.
%Type = Article
\bibitem[{{Peale} and {Boss}(1977)}]{Peale77}
\bibinfo{author}{{Peale}, S.J.}, \bibinfo{author}{{Boss}, A.P.},
  \bibinfo{year}{1977}.
\newblock \bibinfo{title}{{A spin-orbit constraint on the viscosity of a
  Mercurian liquid core}}.
\newblock \bibinfo{journal}{J. Geophys. Res.} \bibinfo{volume}{82},
  \bibinfo{pages}{743--749}.
%Type = Article
\bibitem[{{Peale} and {Cassen}(1978)}]{Peale78}
\bibinfo{author}{{Peale}, S.J.}, \bibinfo{author}{{Cassen}, P.},
  \bibinfo{year}{1978}.
\newblock \bibinfo{title}{{Contribution of tidal dissipation to lunar thermal
  history}}.
\newblock \bibinfo{journal}{Icarus} \bibinfo{volume}{36},
  \bibinfo{pages}{245--269}.
%Type = Article
\bibitem[{{Peale} et~al.(1980){Peale}, {Cassen} and {Reynolds}}]{Peale80}
\bibinfo{author}{{Peale}, S.J.}, \bibinfo{author}{{Cassen}, P.},
  \bibinfo{author}{{Reynolds}, R.T.}, \bibinfo{year}{1980}.
\newblock \bibinfo{title}{{Tidal dissipation, orbital evolution, and the nature
  of Saturn's inner satellites}}.
\newblock \bibinfo{journal}{Icarus} \bibinfo{volume}{43},
  \bibinfo{pages}{65--72}.
%Type = Article
\bibitem[{{Person} et~al.(2006){Person}, {Elliot}, {Gulbis}, {Pasachoff},
  {Babcock}, {Souza} and {Gangestad}}]{Person06}
\bibinfo{author}{{Person}, M.J.}, \bibinfo{author}{{Elliot}, J.L.},
  \bibinfo{author}{{Gulbis}, A.A.S.}, \bibinfo{author}{{Pasachoff}, J.M.},
  \bibinfo{author}{{Babcock}, B.A.}, \bibinfo{author}{{Souza}, S.P.},
  \bibinfo{author}{{Gangestad}, J.}, \bibinfo{year}{2006}.
\newblock \bibinfo{title}{{Charon's radius and density from the combined data
  sets of the 2005 July 11 occultation}}.
\newblock \bibinfo{journal}{Astron. J.} \bibinfo{volume}{132},
  \bibinfo{pages}{1575--1580}.
%Type = Book
\bibitem[{{Press} et~al.(1992){Press}, {Teukolsky}, {Vetterling} and
  {Flannery}}]{Press92}
\bibinfo{author}{{Press}, W.H.}, \bibinfo{author}{{Teukolsky}, S.A.},
  \bibinfo{author}{{Vetterling}, W.T.}, \bibinfo{author}{{Flannery}, B.P.},
  \bibinfo{year}{1992}.
\newblock \bibinfo{title}{{Numerical Recipes in Fortran 77. The Art of
  Scientific Computing}}.
\newblock \bibinfo{publisher}{Cambridge Univ. Press},
  \bibinfo{address}{Cambridge, Ch.~16}.
%Type = Article
\bibitem[{{Rauch} and {Holman}(1999)}]{Rauch99}
\bibinfo{author}{{Rauch}, K.P.}, \bibinfo{author}{{Holman}, M.},
  \bibinfo{year}{1999}.
\newblock \bibinfo{title}{{Dynamical chaos in the Wisdom-Holman integrator:
  Origins and solutions}}.
\newblock \bibinfo{journal}{Astron. J.} \bibinfo{volume}{117},
  \bibinfo{pages}{1087--1102}.
%Type = Article
\bibitem[{{Rodr{\'{\i}}guez} et~al.(2012){Rodr{\'{\i}}guez}, {Callegari},
  {Michtchenko} and {Hussmann}}]{Rodriguez12}
\bibinfo{author}{{Rodr{\'{\i}}guez}, A.}, \bibinfo{author}{{Callegari}, N.},
  \bibinfo{author}{{Michtchenko}, T.A.}, \bibinfo{author}{{Hussmann}, H.},
  \bibinfo{year}{2012}.
\newblock \bibinfo{title}{{Spin-orbit coupling for tidally evolving
  super-Earths}}.
\newblock \bibinfo{journal}{Mon. Not. R. Astron. Soc.} \bibinfo{volume}{427},
  \bibinfo{pages}{2239--2250}.
%Type = Article
\bibitem[{{Singer}(1968)}]{Singer68}
\bibinfo{author}{{Singer}, S.F.}, \bibinfo{year}{1968}.
\newblock \bibinfo{title}{{The origin of the Moon and geophysical
  consequences}}.
\newblock \bibinfo{journal}{GJRAS} \bibinfo{volume}{15},
  \bibinfo{pages}{205--226}.
%Type = Article
\bibitem[{{Stern}(1992)}]{Stern92}
\bibinfo{author}{{Stern}, S.A.}, \bibinfo{year}{1992}.
\newblock \bibinfo{title}{{The Pluto-Charon system}}.
\newblock \bibinfo{journal}{Ann. Rev. Astron. Astrophys.} \bibinfo{volume}{30},
  \bibinfo{pages}{185--233}.
%Type = Article
\bibitem[{{Taylor} and {Margot}(2010)}]{Taylor10}
\bibinfo{author}{{Taylor}, P.A.}, \bibinfo{author}{{Margot}, J.L.},
  \bibinfo{year}{2010}.
\newblock \bibinfo{title}{{Tidal evolution of close binary asteroid systems}}.
\newblock \bibinfo{journal}{Celest. Mech. Dyn. Astron.} \bibinfo{volume}{108},
  \bibinfo{pages}{315--338}.
%Type = Article
\bibitem[{{Taylor} and {Margot}(2011)}]{Taylor11}
\bibinfo{author}{{Taylor}, P.A.}, \bibinfo{author}{{Margot}, J.L.},
  \bibinfo{year}{2011}.
\newblock \bibinfo{title}{{Binary asteroid systems: Tidal end states and
  estimates of material properties}}.
\newblock \bibinfo{journal}{Icarus} \bibinfo{volume}{212},
  \bibinfo{pages}{661--676}.
%Type = Article
\bibitem[{{Tholen} et~al.(2008){Tholen}, {Buie}, {Grundy} and
  {Elliott}}]{Tholen08}
\bibinfo{author}{{Tholen}, D.J.}, \bibinfo{author}{{Buie}, M.W.},
  \bibinfo{author}{{Grundy}, W.M.}, \bibinfo{author}{{Elliott}, G.T.},
  \bibinfo{year}{2008}.
\newblock \bibinfo{title}{{Masses of Nix and Hydra}}.
\newblock \bibinfo{journal}{Astron. J.} \bibinfo{volume}{135},
  \bibinfo{pages}{777--784}.
%Type = Article
\bibitem[{{Touma} and {Wisdom}(1994a)}]{Touma94a}
\bibinfo{author}{{Touma}, J.}, \bibinfo{author}{{Wisdom}, J.},
  \bibinfo{year}{1994}a.
\newblock \bibinfo{title}{{Evolution of the Earth-Moon system}}.
\newblock \bibinfo{journal}{Astron. J.} \bibinfo{volume}{108},
  \bibinfo{pages}{1943--1961}.
%Type = Article
\bibitem[{{Touma} and {Wisdom}(1994b)}]{Touma94b}
\bibinfo{author}{{Touma}, J.}, \bibinfo{author}{{Wisdom}, J.},
  \bibinfo{year}{1994}b.
\newblock \bibinfo{title}{{Lie-Poisson integrators for rigid body dynamics in
  the solar system}}.
\newblock \bibinfo{journal}{Astron. J.} \bibinfo{volume}{107},
  \bibinfo{pages}{1189--1202}.
%Type = Article
\bibitem[{{Touma} and {Wisdom}(1998)}]{Touma98}
\bibinfo{author}{{Touma}, J.}, \bibinfo{author}{{Wisdom}, J.},
  \bibinfo{year}{1998}.
\newblock \bibinfo{title}{{Resonances in the early evolution of the Earth-Moon
  system}}.
\newblock \bibinfo{journal}{Astron. J.} \bibinfo{volume}{115},
  \bibinfo{pages}{1653--1663}.
%Type = Article
\bibitem[{{Ward} and {Canup}(2006)}]{Ward06}
\bibinfo{author}{{Ward}, W.R.}, \bibinfo{author}{{Canup}, R.M.},
  \bibinfo{year}{2006}.
\newblock \bibinfo{title}{{Forced resonant migration of Pluto's outer
  satellites by Charon}}.
\newblock \bibinfo{journal}{Science} \bibinfo{volume}{313},
  \bibinfo{pages}{1107}.
%Type = Article
\bibitem[{{Wisdom} and {Holman}(1991)}]{Wisdom91}
\bibinfo{author}{{Wisdom}, J.}, \bibinfo{author}{{Holman}, M.},
  \bibinfo{year}{1991}.
\newblock \bibinfo{title}{{Symplectic maps for the n-body problem}}.
\newblock \bibinfo{journal}{Astron. J.} \bibinfo{volume}{102},
  \bibinfo{pages}{1528--1538}.
%Type = Article
\bibitem[{{Yoder} and {Peale}(1981)}]{Yoder81}
\bibinfo{author}{{Yoder}, C.F.}, \bibinfo{author}{{Peale}, S.J.},
  \bibinfo{year}{1981}.
\newblock \bibinfo{title}{{The tides of Io}}.
\newblock \bibinfo{journal}{Icarus} \bibinfo{volume}{47},
  \bibinfo{pages}{1--35}.
%Type = Article
\bibitem[{{Zahn}(1977)}]{Zahn77}
\bibinfo{author}{{Zahn}, J.P.}, \bibinfo{year}{1977}.
\newblock \bibinfo{title}{{Tidal friction in close binary stars}}.
\newblock \bibinfo{journal}{Astron. Astrophys.} \bibinfo{volume}{57},
  \bibinfo{pages}{383--394}.

\end{thebibliography}

\end{document}